\documentclass[aip,jcp, a4paper]{revtex4-1}
    \usepackage[english]{babel}
    \usepackage{lmodern}
    \usepackage{amsmath}
    \usepackage{amssymb} 
    \usepackage{tabularx} 
    \usepackage{amsfonts}
    \usepackage[usenames,dvipsnames]{pstricks}
    \usepackage{multirow}    
    \usepackage{braket}
    \usepackage[caption=false]{subfig} 
    \usepackage{epsfig}
    \usepackage[]{xcolor}
    \usepackage[]{color}
    \usepackage[english]{babel}
    \usepackage[]{colortbl}
    \usepackage{pst-grad} % For gradients
    \usepackage{pst-plot} % For axes
    \usepackage[colorlinks,hyperindex]{hyperref}
    \usepackage[sort&compress]{natbib}
        
    \hypersetup
    {
        colorlinks,%
        citecolor=black,%
        linkcolor=black,%
        urlcolor=black,%
    }

    \newcommand{\comut}[2]{\left[ #1 , #2 \right]} % THE COMUTATOR

    \DeclareMathOperator{\Tr}{Tr}
 
\setcitestyle{super,open={},close={}}

\makeindex
\begin{document}

\title{Controlling the conductance of molecular junctions using proton transfer reactions: A theoretical model study}
\author{Chriszandro Hofmeister, Pedro B.\ Coto, and Michael Thoss}

%\email{chriszandro.hofmeister@fau.de}
%\homepage{http://www.thcp.physik.uni-erlangen.de}
\affiliation{Institut f{\"u}r Theoretische Physik und Interdisziplin{\"a}res Zentrum f{\"u}r Molekulare Materialien \\
Friedrich-Alexander-Universit{\"a}t Erlangen-N{\"u}rnberg \\ 
Staudtstr.\ 7/B2, D-91058 Erlangen, Germany }
%\keywords{molecular junction, proton transfer, molecular switch, master %equation, double well}

%\date{today}

\begin{abstract}
The influence of an intramolecular proton transfer reaction on the conductance of a molecular junction is investigated employing a generic model, which includes the effects of the electric field of the gate and leads electrodes and the coupling to a dissipative environment. Using a quantum master equation approach it is shown that, depending on the localization of the proton, the junction exhibits a high or low current state, which can be controlled by external electric fields. Considering different regimes, which range from weak to strong hydrogen bonds in the proton transfer complex and comprise situations with high and low barriers, necessary preconditions to achieve control are analyzed. The results show that systems with a weak hydrogen bond and a significant energy barrier for the proton transfer can be used as molecular transistors or diodes.  
\end{abstract}

\maketitle

\section{Introduction} 
The field of molecular-scale electronics has seen tremendous progress in recent years, both with respect to experimental investigations and theoretical studies of the underlying transport mechanisms.\cite{Cuniberti05,Cuevas2010,Zimbovskaya2011,Bergfield13,Baldea15,Xiang16} The most widely studied architecture in this field is a molecular junction, where a single molecule is bound to metal or semiconductor electrodes. Molecular junctions provide interesting systems to study basic mechanisms of non-equilibrium charge transport in a many-body quantum system at the nanoscale. An intriguing question concerns the possibility to realize the functionality of an electronic device with a single molecule in a molecular junction. The theoretical proposal of a molecular rectifier several decades ago \cite{Aviram74} can be considered as a starting point of the field. In recent years, experimental studies have shown that the current-voltage
characteristics of molecular junctions may resemble those of basic electronic devices, such as
rectifiers \cite{Elbing2005,Loertscher2012,Capozzi2015} or transistors. \cite{Park2000,Perrin15} An important element for the design of molecular memory or logic
devices is a molecular switch. \cite{vanderMolen10} A molecular junction may be used as a nanoswitch, if the molecule
can exist in two or more differently conducting states that are sufficiently stable and can be reversibly
transferred into each other.

A variety of different mechanisms have been proposed 
to achieve reversible switching of  molecular junctions
between different conductance states.
\cite{Donhauser01,Moresco01,Dulic03,LiSankey04,Mendes05,Choi06,delValle07,Liljeroth07,vanderMolen10} 
Most mechanism for optical switches considered so far are based on light-induced 
conformational changes, in particular 
isomerization reactions, 
\cite{Choi06,delValle07}
or ring-opening reactions  \cite{Dulic03,LiSankey04} 
of the molecular bridge. Nonoptical mechanisms include
reversible redox reactions, 
for example, in catenane and
rotaxane molecules triggered by  voltage pulses. \cite{Mendes05}

As an alternative mechanism for switching of molecular junctions, hydrogen tautomerization or proton transfer reactions
have been proposed  recently. \cite{Liljeroth07,Benesch09,Pan09,Auwaerter11}
Employing STM experiments, it was demonstrated that naphtalocyanine molecules
at Cu(110) \cite{Liljeroth07} and porphyrin at Ag(111) \cite{Auwaerter11} show current-induced switching,
which is caused by hydrogen transfer. 
Moreover, photoinduced excited state hydrogen transfer was proposed as a mechanism to switch molecular junctions. \cite{Benesch09} 
Simulations show that this mechanism is also active in molecular junctions
that use carbon nanotubes  as electrodes. \cite{Zhao10}
In contrast to  other mechanisms suggested previously to realize molecular switches, hydrogen (or proton) translocation within the 
molecular bridge has the advantage that the overall length and thus the molecule-electrode 
binding geometry of the junction is not changed significantly. This makes this
mechanism a promising candidate for a molecular switch.

\begin{figure}[htbp] 
\begin{center} 
\includegraphics[width=0.5\textwidth]{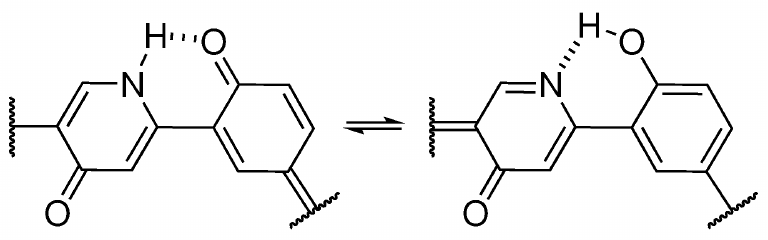} 
\caption{Scheme of an intramolecular proton
        transfer reaction triggered by an external electrostatic field. The tautomers exhibit distinctive conductance properties as a consequence of their different electronic structure.} \label{fig:tautomerization} 
\end{center} 
\end{figure}

In previous work,\cite{hofmeister_switching_2014} we have shown that a proton transfer reaction triggered by an external electrostatic field provides another possibility to control the conductance state of a molecular junction. A prototype example is the reaction depicted in Fig.\ \ref{fig:tautomerization}, which involves proton transfer between a nitrogen and an oxygen center in a hydrogen-bonding complex. Similar results have been reported also for different systems.\cite{jankowska_electric_2015}

So far, our studies of proton transfer in molecular junctions considered the conductance for nuclei fixed at the geometry of the enol or keto tautomer. A complete description requires to incorporate the coupling of the electrons to the motion of the proton in the junction at finite bias voltage, which is a challenging nonequilibrium transport problem. The theoretical description is further complicated by the fact that the large amplitude motion of the proton requires a potential beyond the harmonic approximation. As a consequence, theoretical approaches which rely on the harmonic approximation for the vibrational degrees of freedom cannot be used. In the present paper, we use a generic model with parameters motivated by our previous first-principles studies\cite{hofmeister_switching_2014} and employ a quantum master equation at the level of Redfield theory to tackle this problem. Similar approaches, sometimes combined with classical approximations or inelastic scattering theory, have been used before to investigate transport in molecular junctions involving large amplitude motion.\cite{Cizek05,Koch06b,Donarini06,Pshenichnyuk11,Kecke12,Dzhioev13,Pozner14} Based on this methodology, we analyze the transport properties and show that the conductance state of a molecular junction can indeed be controlled using a proton transfer reaction. Thereby, we consider both the influence of gate and bias voltages on the motion of the proton. Depending on the specific realization and parameters, the current-voltage characteristics of the junction can resemble those of a diode or a transistor.

The paper is organized as follows: After an introduction of the model in Sec.\ \ref{sec:modell}, we outline the theoretical methodology in Sec \ref{sec:theory}. The results are presented in Sec. \ref{sec:results}, divided into two parts. In the first part, we consider a system with a weak hydrogen bond in the proton transfer complex corresponding to a high barrier. In the second part, we discuss results for systems with stronger hydrogen bonding. Sec.\ \ref{sec:conclusion} summarizes and concludes.

\section{Model\label{sec:modell} } 

To study the influence of an intramolecular  proton transfer process on the conductance of a molecular junction we consider a model described by the Hamiltonian 
\begin{equation}\label{eq:hamilton} 
H = H_{S} + H_{L} + H_{B} + H_{SL} + H_{SB}, 
\end{equation}
where $H_S$ corresponds to the molecule (in the following also referred to as the system), $H_{L}$ to the left and right leads and $H_{B}$ represents a dissipative environment. The coupling of the molecule to the leads and the environment are denoted by  $H_{SL}$ and
$H_{SB}$, respectively, and taken together define the system-reservoir coupling $H_{SR} = H_{SL} + H_{SB}$.

%---------------------------------------------------------------
The Hamiltonian of the system, $H_S$, describes not only the electronic states of the molecule but it also accounts for the motion of the proton. In our model, we consider a single electronic state of the molecule (the bridge state) in the basis defined by the neutral and charged state of the molecule, ($\ket{0}$) and ($\ket{1}$), respectively.  Denoting the corresponding Hamiltonian elements by $h_0$ and $h_1$, respectively, $H_S$ can be expressed as
\begin{equation} 
 H_S= h_0 dd^\dagger + h_1 d^\dagger d , \label{eq:system-hamiltonian} 
\end{equation} 
where $d^\dagger$ and $d$ are the fermionic creation and annihilation operators. The Hamiltonian elements take the form
\begin{equation} 
 h_i = -\frac{1}{2M} \frac{\partial^2}{\partial x^2}+ V_i(x),
\end{equation}
where $M$ denotes the mass of the proton. 

\begin{figure}[htb]
	\begin{center} \includegraphics[width=0.5\textwidth]{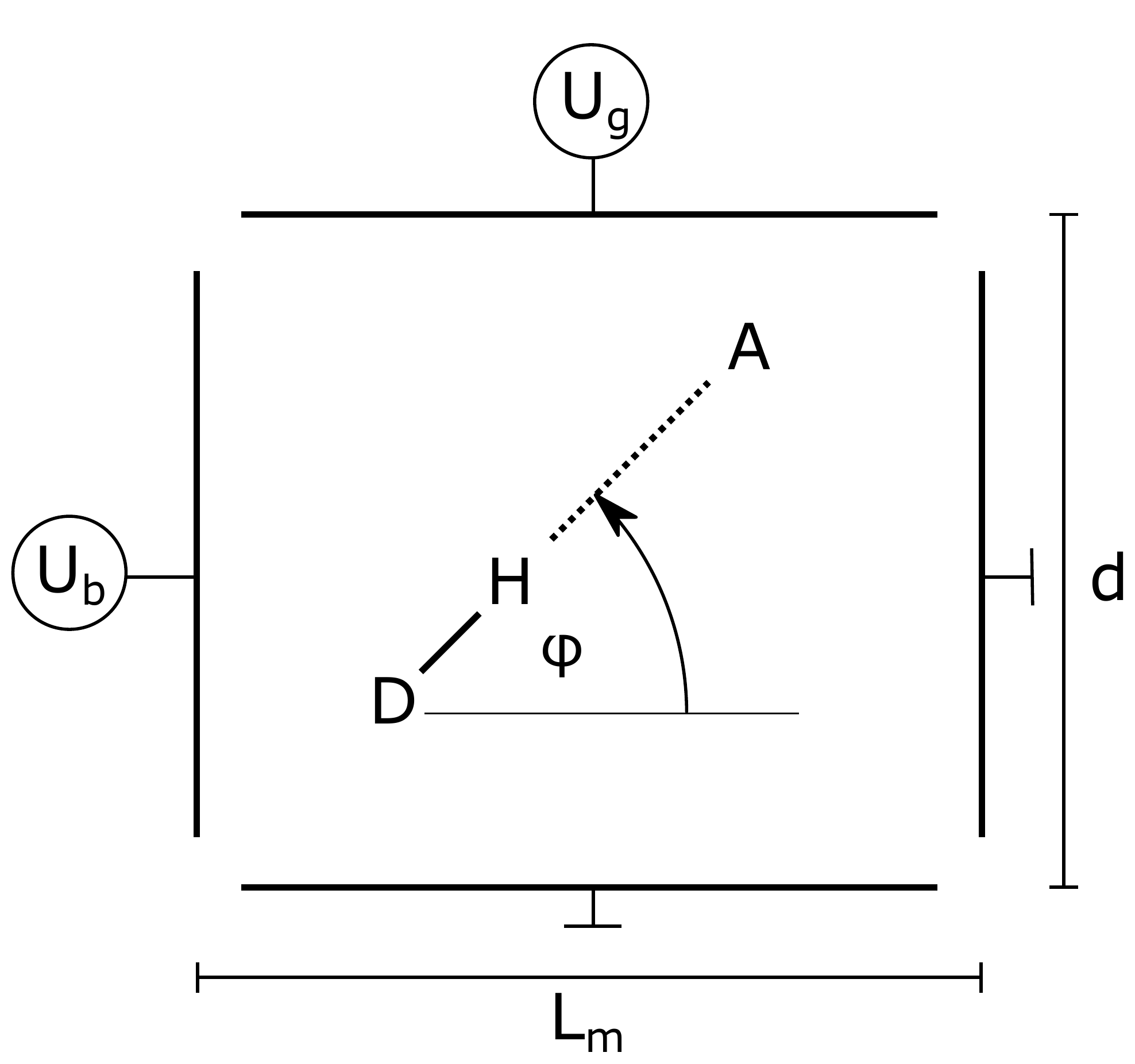}
		\caption{Schematic representation of the proton transfer model system investigated in this work. The angle between the transfer path and the bias field tuned by $U_{b}$ is denoted with $\phi$. The bias field is perpendicular to the gate field tuned by $U_{g}$.}
		\label{fig:angle} 
	\end{center} \end{figure}

\begin{figure}[htpb]
	\includegraphics[width=0.5\textwidth]{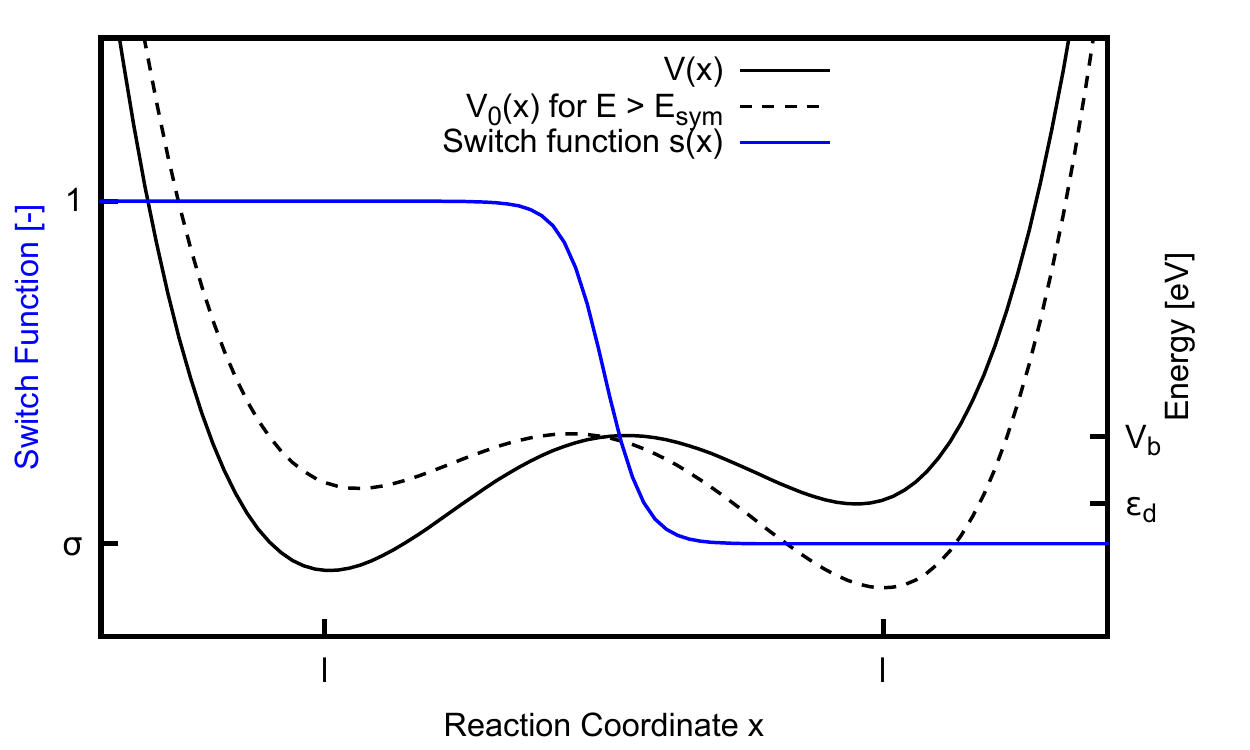}
	\caption{Potentials used to model the intramolecular proton transfer reaction with the parameters barrier energy $V_{b}$, detuning energy $\epsilon_{d}$,  and distance of the minima $2l$ defining the translocation length $x_{T}$. $V(x)$ is the potential without external fields, while $V_0(x)$ includes an external field of the gate electrode (shown for a value of $E > E_{sym}$). The switch
		function $s(x)$ scales the molecule-lead coupling  depending on the	location of the proton.}.\label{fig:potentials}
	\end{figure}

The potential of the neutral state is given by $V_{0}(x) = V(x) + V_{\rm ext}(x)$ where $V(x)$ is the potential for the intramolecular motion of the proton and $V_{\rm ext}(x)$ accounts for the influence of external fields, in the present case the gate and the leads electric potentials.
To describe the motion of the proton between the donor (D) and acceptor (A) moieties (cf.\ Fig.\ \ref{fig:angle}), we use a double-well potential given by 
\begin{equation} \label{double_well}
V(x) = \frac{\varepsilon_{d}}{2l}(x+l) + \frac{V_b - 0.5 \epsilon_{d}}{l^4} (x+l)^2(x-l)^2, 
\end{equation} 
where the minima of the potential are separated by a distance $x_{T} = 2l$ and the energy of the right
well is detuned with respect to that of the left by $\epsilon_{d}$ to account for the general non-symmetrical case where the donor and acceptor states have different stabilities.\cite{hofmeister_switching_2014} In Eq.~(\ref{double_well}), $V_b=V(0)$ denotes the potential energy at $x=0$, i.e.\ the maximum of the double-well potential, and is in the
following referred to as the barrier energy of the proton transfer process (see Fig.\ \ref{fig:potentials} for a schematic representation). In general, this potential may be different in the neutral and charged state of the molecule. Here we assume for
simplicity that upon charging, the potential only acquires a constant shift given by the electron affinity $\epsilon_{0}$ yielding $V_1(x) =
V_0(x) + \epsilon_0$ for the potential of the charged state. In the calculations reported below a value of $\epsilon_0=0.1$ eV is chosen. 
The effect of changes of the form of the potential upon charging on the transport properties will be considered in future work.

The distances ($2l$) between the minima of the double-well potential employed in the simulations have been selected to characterize different hydrogen bond situations. Employing the acceptor (A) proton distance in the donor state, $d_{D}(AH)$, a hydrogen bond is considered weak for $d_{D}(AH) > 2.0$ {\AA} and strong for $d_{D}(AH) < 1.5$
\AA.\cite{Desiraju_weak_2001,mckenzie_effect_2014,perrin_strong_1997} Accordingly, in this work we have considered three sets
of parameters (see Table \ref{tab:parameters}) corresponding to weak, medium and strong hydrogen bonding regimes. In addition, for simplicity, the equilibrium bond distances $d(DH)$ and $d(AH)$ in both the donor and acceptor states have been taken as $d(DH)=d(AH)=1.0$ \AA.     
Using these values, the translocation length of the proton is given by the relation $x_{T} = d_{D}(AH) - d(DH)$.  
An aspect that strongly affects the transport properties is the height of the barrier of the double-well potential. In this work, we have considered the situation where the system with weak hydrogen bonding exhibits a significant barrier while the other two systems have rather small or vanishing barriers.

The effects of the external field of the gate and
source/drain electrodes on the motion of the proton depend
in principle on the specific nature of the chemical species involved in the process. To account for these effects in our simulations we use a simplified model that is based on the direction of the transfer path of the proton between the donor and acceptor moieties (see Fig.\ \ref{fig:angle}). The direction determined by the angle $\phi$ characterizes the influence of the electric fields of the gate and the lead electrodes (which are assumed to be perpendicular) and is described by the external electrostatic potential $V_{\rm ext}=-E x$ with
\begin{equation} 
E = -\left( \frac{U_{g} }{d} \sin(\phi)  + \frac{U_{b}}{L_m} \cos(\phi)  \right), \label{eq:external_field} 
\end{equation}
where $U_{b}$ and $U_{g}$ are the bias and the gate voltage, respectively, $L_{m}$ is the length of the molecular bridge, and $d$ is the distance between the gate electrode and the junction.  In the simulations reported below, we have used $L_{m}=d=1$ nm. 

In our model, the left (l) and right (r) leads are described by reservoirs of noninteracting electrons, 
\begin{equation} H_{L} = \sum_{\alpha \in r,l}\sum_{k_\alpha}
    \epsilon_{k_\alpha} c_{k_\alpha}^\dagger c_{k_\alpha}, 
\end{equation}
where  $c_{k_\alpha}$ and $c_{k_\alpha}^\dagger$ denote the creation 
and annihilation operators for
an electron with energy $\epsilon_{k_\alpha}$ in the $k$-th state of lead $\alpha$.
In equilibrium, the occupation of the lead states is given by the Fermi
distribution 
$f_{\alpha}(\varepsilon) = 1 \mathbin{/} \left( 1 - \exp \left[ \beta \left( {\varepsilon - \mu_{\alpha}}  \right] \right) \right)$ 
where
$\beta = 1 \mathbin{/} k_{B}T$ 
and $\mu_{\alpha}$ is the chemical potential of the lead. 
The temperature of the reservoirs is set to $T=293$ K for all calculations presented in this work.
For finite bias voltage $U_{b}$, we assume a symmetric shift of the chemical potentials, $\mu_{l,r} = \pm \frac{U_{b}}{2}$, with respect
to the Fermi level at $\epsilon_f=0$ eV.  

The molecule-lead interaction is described by 
\begin{equation} H_{SL} = \sum_{\alpha \in l,r} \sum_{k_\alpha}
	V_{k_\alpha}s(x)(d^\dagger c_{k_\alpha} + c_{k_\alpha}^\dagger d).
	\label{eq:molecule-lead} 
\end{equation} 
Thereby, the molecule-lead coupling strengths $V_{k_{\alpha}}$ and the electronic energies
$\epsilon_{k_{\alpha}}$ are defined by the level-width function $\Gamma_{\alpha}(E) = 2 \pi \sum_{k_{\alpha}} |V_{k_{\alpha}}|^2
\delta(E-\epsilon_{k_{\alpha}}).$ In this work, we consider the leads to be semi-infinite chains described by a tight-binding model,
\cite{Cizek04,peskin_introduction_2010}
which implies \begin{equation} \label{eq:franckSS} \Gamma_\alpha(E) = \begin{cases} \frac{\nu^2}{\gamma^2}\sqrt{ 4\gamma^2 - (E -
		\mu_\alpha)^2} & |E| \geq 2 |\gamma| \\ 0 & |E| < 2 |\gamma| \end{cases} \end{equation} The parameters $\gamma$ and $\nu$ in
$\Gamma_{\alpha}(E)$, describing the coupling strength between two neighboring sites in the leads and between the molecule and the leads,
respectively, are set to $\gamma=3$ eV and $\nu=0.1$ eV corresponding to a weak molecule-lead coupling and a band width of 6 eV.

To model the dependence of the conductance on the position of the proton, we introduce a dimensionless switch function 
\begin{equation}\label{eq:switch_function} 
s(x) = \frac{1+\sigma}{2} - \frac{1-\sigma}{2} \tanh(x), 
\end{equation}
which effectively reduces the coupling matrix elements $V_{k_\alpha}$ by
the factor $\sigma$ when the proton is localized in the right potential well. In the calculations presented below, a value of $\sigma = 0.1$ is used.  As a result (see below), the conductance is larger (``ON'' state) for a proton located in the left well and smaller (``OFF'' state) in the right well. In addition, depending on the strength of the external electric field $E$, the global minimum of the potential may be tuned from the left to the right well, as illustrated in Fig. \ref{fig:potentials}.

In a molecular junction the motion of the proton is coupled to other vibrational degrees of freedom of the molecular bridge as well as to phonons in the leads, which may cause energy relaxation. To model this effect, we consider the coupling of the proton to a reservoir of harmonic oscillators, 
\begin{equation} 
H_{B}  = \sum_{j} \omega_{j} b_j^{\dagger} b_j , \label{eq:harmonic_bath} 
\end{equation} 
where $b_j^\dagger$ and $b_j$ are the creation and annihilation operators of the harmonic oscillator with frequency $\omega_j$ and assume that the proton couples linearly to the oscillators of the reservoir, 
\begin{equation} H_{SB} = \sum_{j} \lambda_{j} q_{j} x
    \label{eqn:coupling_system-bosons},
\end{equation} 
where $q_{j} = (b_{j} +
b_{j}^{\dagger})/\sqrt{2}$ denotes the coordinate of the $j$th bath oscillator. The individual coupling strengths $\lambda_{j}$ are characterized by the bath
spectral density given by $J(\omega)=\sum_{j}\lambda_j^{2}\delta(\omega-\omega_j)$. 
In this work, we use an Ohmic bath with spectral density
\begin{equation} 
J(\omega) = \eta \omega e^{-\frac{\omega}{\omega_{c}}}. 
\label{eqn:spectral-density-ohmic}
\end{equation}
The maximum of the spectral density is given by the
characteristic frequency, $\omega_{c}= 0.097$ eV, and the parameter $\eta$ defines the overall coupling strength (see below).

\begin{table}[htb] \begin{tabular}{|l|c|c|c|} \hline System & \textbf{Weak}  & \textbf{Medium} & \textbf{Strong} \\ \hline
        $d_{D}(AH)$ (\AA)  & 2.5 & 2.0 & 1.5 \\ \hline 
        $l$ (\AA) & 0.75  & 0.50 & 0.25 \\ \hline 
        $V_{b}$ (eV) & 0.8  &  0.2  & 0.025 \\ \hline
        $\epsilon_{d}$ (eV) & 0.3  &  0.1  & 0.05 \\ \hline 
      %  $U_{sym}$ in eV & 2.0 & 1.0  & 0.5  
      %  \\ \hline
 \end{tabular}
    \caption{Parameter sets for the three systems considered corresponding to weak, medium, and strong hydrogen bonding regimes.}
    \label{tab:parameters} 
\end{table}

\section{Methodology} \label{sec:theory} 

We treat the molecular junction in the framework of open quantum systems, where the molecule is the
system of interest.  The reduced density matrix of the molecule, $\rho$, is defined by tracing over the reservoir degrees
of freedom, $\rho = \Tr_{R}(\varrho)$, where $\varrho$ denotes the density matrix of the overall system. The dynamics of the reduced density
matrix is described employing a Markovian quantum master equation in the weak coupling limit, also known as Redfield equations
\cite{timm_tunneling_2008-1,Nitzan_book} given by 
\begin{equation}\label{eq:redfield}
    \frac{\partial \rho}{\partial t}=
\mathcal{L}\rho(t) = \mathcal{L}_S\rho(t) + \mathcal{R}\rho(t).  \end{equation} 
Thereby, the Liouvillian superoperators read 
\begin{eqnarray}
\mathcal{R}\rho(t) &=& -\Tr_{R} \int_{0}^{\infty} d\tau \comut{H_{SR}}{\comut{H_{SR}(-\tau)} {\rho(t)\otimes\rho_R}}, \nonumber \\
\mathcal{L}_S\rho(t) &=& -i\comut{H_{S}}{\rho(t)},  
\end{eqnarray} 
where $H_{SR}(\tau) = e^{-i(H_S + H_{L} + H_B)\tau}H_{SR}e^{i(H_S
+ H_{L} + H_B)\tau}$ is the molecule-reservoir interaction 
%$H_{SR} = H_{LS} + H_{SB}$ 
transformed to the interaction picture.

Employing the eigenstates of $H_{S}$, $\left\{ \ket{m, i} \right\}$, where $\ket{m, 0}$ denotes the vibrational states of the neutral
molecule, i.e.\ the eigenstates of $h_0$, and $\ket{m, 1}$ those of the charged molecule ($h_1$),  Eq.\ (\ref{eq:redfield}) reads
\begin{equation} \dot\rho_{mn}^{jk} = \sum_{l, \mu \nu} {\mathcal{L}_S}_{mn, \mu \nu}^{jl}  \rho_{\mu \nu}^{lk} + \mathcal{R}_{mn, \mu
	\nu}^{jl}  \rho_{\mu \nu}^{lk}, \label{eq:explicit-redfield} \end{equation} where we use the notation $\rho_{mn}^{jk} = \bra{m,j} \rho
\ket{n,k}$.  
%Free Livioulion
The operator $\mathcal{L}_S$ governs the free time evolution of the molecule in absence of the reservoirs and is given by 
\begin{subequations}
\begin{eqnarray}
{\mathcal{L}_S^{00}}_{n_1^\prime n_2^\prime, n_1 n_2} &=& i \delta_{n_2^\prime n_2}\delta_{n_1^\prime n_1}(E_{n_2}^0 -
		E_{n_1}^0) \\ {\mathcal{L}_S^{11}}_{v_1^\prime v_2^\prime, v_1 v_2} &=& i \delta_{v_2^\prime v_2}\delta_{v_1^\prime
		v_1}(E_{v_2}^1-E_{v_1}^1).  
\end{eqnarray}
\end{subequations}

The relaxation operator $\mathcal{R}$ describes the interaction with the
reservoirs including the leads and the harmonic bath, $\mathcal{R} =
\mathcal{R}_{L} + \mathcal{R}_{B}$.  Thereby, the initial state of the
reservoir is assumed to be a product state $\rho_{R} = \rho_{L}\rho_{B}$ with
the initial density matrix of the leads 
\begin{equation} \rho_{L} = \exp(-\beta
	H_{L} - \sum_{\alpha \in \left\{ l,r \right\}} \mu_{\alpha} \sum_{k_\alpha }
	c_{k_\alpha}^{\dagger}c_{k_\alpha}) \label{eq:equil_leads}
\end{equation}
and for the harmonic bath $\rho_{B} = \exp(-\beta H_{B}) $. In the $\left\{ \ket{m, i} \right\}$ basis the relaxation
operator reads 
\begin{widetext} 
\begin{subequations}
\begin{eqnarray}
			\label{eq:explicit-redfield-energy}
			{\mathcal{R}^{00}_{\alpha}}_{n_1^\prime n_2^\prime,n_1 n_2} &=&
			-\frac{1}{2}\delta_{n_1 n_1^\prime} \sum_{v}
			\Lambda_{+}^{\alpha}(\omega_{n_2 v}) S_{n_2 v} S_{ n_2^\prime v} -
			\frac{1}{2} \delta_{n_2 n_2^\prime} \sum_{v} \Lambda_{+}^{\alpha}(
			\omega_{ n_1 v }) S_{ n_1 v} S_{  n_1^\prime v} \\
			{\mathcal{R}_{\alpha}^{11}}_{v_1^\prime v_2^\prime, v_1 v_2} &=&
			-\frac{1}{2} \delta_{v_1 v_1^\prime}\sum_{n} \Lambda_{-}^{\alpha}(
			\omega_{n v_2} ) S_{n v_2} S_{n v_{2}^\prime } - \frac{1}{2} \delta_{v_2
			v_2^\prime} \sum_{n} \Lambda_{-}^{\alpha}( \omega_{n v_1}) S_{n v_1} S_{
			n v_1^\prime} \\ {\mathcal{R}_{\alpha}^{01}}_{n_1^\prime n_2^\prime, v_1
			v_2} &=& \frac{1}{2}  \Lambda_{-}^{\alpha}( \omega_{n_1^\prime v_1})
			S_{n_1 v_1}  S_{n_2^\prime v_2 } + \frac{1}{2}  \Lambda_{-}^{\alpha}(
			\omega_{n_2^\prime v_2}) S_{n_2^\prime v_1 } S_{n_1^\prime v_{1}} \\
			{\mathcal{R}_{\alpha}^{10}}_{v_1^\prime v_2^\prime, n_1 n_2} &=&
			\frac{1}{2} \Lambda_{+}^{\alpha}(\omega_{n_1 v_1^\prime}) S_{ n_1
			v_1^\prime } S_{n_2 v_2^\prime} + \frac{1}{2}
			\Lambda_{+}^{\alpha}(\omega_{n_2 v_2^\prime}) S_{n_1 v_1^\prime } S_{n_2
			v_2^\prime }\\ {\mathcal{R}^{ii}_{B}}_{k_1^\prime k_2^\prime, k_1 k_2} &
			=& \delta_{k_{2}^{\prime}k_{2}} \sum_{k} \gamma_{+} (\omega^{i}_{k_{1}k})
			x_{k k_{1}}^{i} x_{k_{1}^{\prime}k}^{i}  +
			\delta_{k_{1}^{\prime} k_{1}} \sum_{k} \gamma_{-}(\omega^{i}_{k k_{2}})
			x_{k_{2}k}^{i} x_{k k_{2}^{\prime}}^{i}  \\  
			&&+			\gamma_{+}(\omega^{i}_{k_{1} k_{1}^{\prime}} )  x_{k_{1}^{\prime}
			k_{1}}^{i} x_{k_{2} k_{2}^{\prime}}^{i} +
			\gamma_{-}(\omega^{i}_{k_{2}^{\prime} k_{2}})  x_{k_{2}
			k_{2}^{\prime}}^{i}  x_{k_{1} k_{1}^{\prime}}^{i}, \nonumber
	\end{eqnarray}
	\end{subequations}
	 \end{widetext} 
where we have introduced the following notation 
\begin{subequations}
\begin{eqnarray}
		\Lambda_{+}^{\alpha}(\omega_{nv}) &=& \Gamma_{\alpha}(\omega_{nv})
		f_{\alpha}(\omega_{nv}), \\ \Lambda_{-}^{\alpha}(\omega_{nv}) &=&
		\Gamma_{\alpha}(\omega_{nv}) \left[1-f_{\alpha}(\omega_{nv}) \right],\\
		\gamma_{+}(\omega_{k^{\prime}k})  &=& 2 \pi J( \omega_{k^{\prime}k})
		\left[1 + n_{B}(\omega_{k^{\prime}k})\right],  \\
		\gamma_{-}(\omega_{k^{\prime}k})  &=& 2 \pi J( \omega_{k^{\prime}k})
		n_{B}\left(\omega_{k^{\prime}k} \right),
	\label{eq:abbreviations-1} 
\end{eqnarray} 
\end{subequations}
with transition frequencies
$\omega_{nv} = E_{v}^1 - E_{n}^0$ and $\omega_{k k^{\prime}}^{i} = E_{k}^i -
E_{k^{\prime}}^i$ as well as transition elements 
$S_{m,v} = \left| \bra{m,0}  s(x)
\ket{v,1} \right|^{2}$ and $x^{i}_{k k^{\prime}} =   \left| \bra{k, i} x
\ket{k^\prime, i}\right|^{2} $.

In the numerical calculations, the system Hamiltonian $H_{S}$ is represented and diagonalized employing a discrete variable representation
(DVR) method optimized for the double-well potential $V(x)$ using Hermite polynomials for the underlying
grid.\cite{littlejohn_general_2002} The stationary state of the reduced density matrix $\tilde{\rho} = \rho(t
\to \infty)$ is obtained by solving the set of linear equations $\mathcal{L}\tilde{\rho} = 0$. The eigenstates employed for the
representation of the density matrix $\rho_{mn}^{ij}$ are truncated at $m,n=60$.  With this truncation, all localized states in the
potential wells and a sufficient number of delocalized states above the energy barrier are included.  To have a proper description of the observables of interest, the coherences of the density matrix have to be included in the calculation. To this end, we include the first six subdiagonals of the density matrix in both occupation spaces. Test calculations show that this provides converged results for the observables of interest. 

The observables of interest comprise the current $I$, the probability distribution for the proton position $\rho(x)$ as well as the average
position $\Braket{x}$. The current is defined as the time derivative of the number of electrons in lead $\alpha$,
\begin{equation}\label{eq:current} 
\hat{I}_{\alpha} = \frac{d}{dt}  c_{k_\alpha}^{\dagger}c_{k_\alpha} = -ie \comut{c_{k_\alpha}^{\dagger}c_{k_\alpha}}{H} 
\end{equation}
which in the $\left\{ \ket{m, i} \right\}$ basis reads 
\begin{eqnarray} 
I_{\alpha} &=&
			\sum_{nv}\Gamma_{\alpha}(\omega_{nv}) S_{nv} \Bigg[ f_{\alpha}(\omega_{nv}) \sum_{n^\prime} S_{n^\prime v} \rho_{n^\prime n}^{00} \nonumber \\
				&& - [ 1 - f_{\alpha}(\omega_{nv})] \sum_{v^\prime} S_{nv^\prime} \rho_{vv^\prime}^{11} \Bigg].
\end{eqnarray}

The probability distribution  $\rho(x_{\beta})$ for the proton position on the DVR grid $\left\{x_{\beta}\right\}$ is given by 
\begin{equation}
\rho(x_{\beta}) = \sum_{i \in \left\{0,1 \right\}}  \sum_{n,m} 
\rho_{nm}^{ii} \phi_{n}^{i} \left( x_{\beta} \right) \phi_{m}^{i} \left( x_{\beta} \right)  \Delta_{\beta},
\label{eq:rho_in_position}
\end{equation}
where $\phi_{n}^{i}\left( x_{\beta} \right)$ are the eigenfunctions of $H_{S}$ in the DVR representation
and $\Delta_{\beta}$ are the corresponding DVR weights. From this, the average position is straightforwardly obtained by the trace formula $ \Braket{x}= \Tr_{S} \left( \hat{x} \rho  \right)$.

We finally comment on the validity of the approximations used in this approach. The treatment of the molecule-reservoir coupling within second order perturbation theory limits the validity of the Redfield equations, Eq.\ (\ref{eq:redfield}), to sufficiently weak coupling between the system and the reservoirs. With respect to the molecule-lead coupling, co-tunneling effects and broadening of features
in the current-voltage characteristics are not included in this approximation. As a result, the validity of the approach requires
temperatures with $k_{B}T  \gg \frac{2 \nu^2 }{\gamma}$, a condition fulfilled in all calculations reported below. With respect to the
coupling of the proton motion to the harmonic bath, the criterion for the validity of the approach is provided by the following condition
for the relaxation tensor elements\cite{egorova_modeling_2003} 
\begin{equation} 
{\mathcal{R}_{B}^{ii}}_{\mu \mu , \nu \nu} \ll   \omega_{\mu \nu }^{i}.
 \label{eq:Validity_two} 
\end{equation} 
This criterion is fulfilled for the parameters considered below.

\section{Results\label{sec:results}} 

The methodology outlined above is in the following applied to analyze the influence of intramolecular proton transfer on the
conductance properties of a molecular junction. For this purpose, we consider the three model systems defined in table \ref{tab:parameters} corresponding to weak, medium and
strong hydrogen bonding.

\subsection{Weakly hydrogen bonded, high-barrier system\label{sec:weak_system_analysis}} 

We first consider the system with weak hydrogen bonding and neglect the influence of the bias voltage on the potential $V_0(x)$ of the
proton motion. This corresponds to a situation where the direction of the translocation path is parallel to the gate field (cf.\ Fig.\
\ref{fig:angle}), i.e.\ $\phi = \pi \mathbin{/} 2$.  The specific form of the double-well potential $V_0(x)$ depends thus only on the gate
voltage $U_{g}$. For the choice $U_{\rm g,sym} = \epsilon_{0} d/(2l)$, the potential $V_0(x)$ is symmetric. This value of the gate
voltage plays a special role (see below) and is in the following referred to as trigger point. For the subsequent discussion, it is useful
to introduce the the detuning voltage $U_{d} = U_{g} - U_{\rm g, sym}$, which is a measure of the asymmetry of the potential $V_{0}(x)$.  The
global minimum of $V_{0}(x)$ is located in the left/right well for $U_{d} \lessgtr 0$.  The influence of the bias voltage on the potential
$V_{0}(x)$ is analyzed in Sec.\ \ref{ref:weak_functionality}.

\subsubsection{Equilibrium properties}

\begin{figure}[hbt]
  \includegraphics[width=0.30\textwidth]{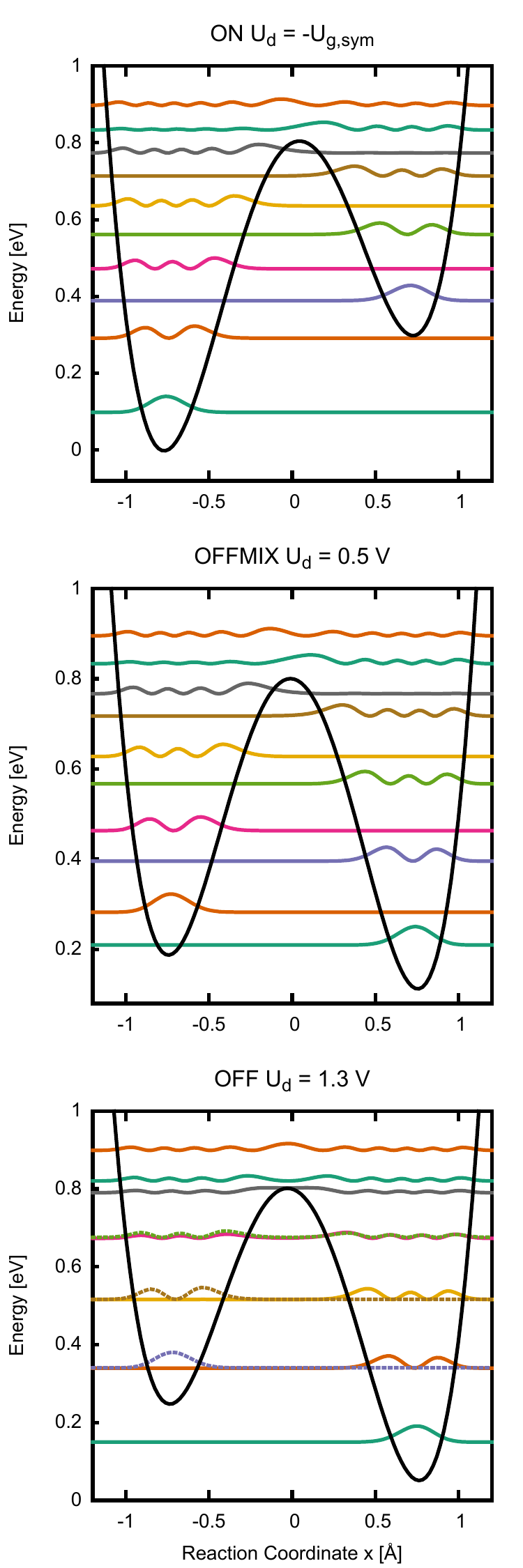}
	\caption{Potentials $V_0(x)$, spectra and density of the eigenfunctions of the weakly bonded system for different detuning voltages $U_{d}$.}
\label{fig:weak_potentials} 
\end{figure}

To facilitate the analysis of the transport properties in Sec.\
\ref{sec:biasing_weak}, we first discuss the equilibrium properties. Fig.\ \ref{fig:weak_potentials} shows the potential $V_0(x)$ for three different
choices of the detuning voltage, $U_{d} =-U_{\rm g,sym} $ (ON), $U_{d} = 0.5$ V
(OFFMIX) and $U_{d} = 1.3$ V (OFF), where the labels in parenthesis are related
to the transport properties (see below), and the respective eigenenergies and eigenstates.  The potentials for the ON/OFF cases
exhibit a pronounced asymmetry with a global minimum in the left/right well,
while the case OFFMIX represents a potential that is closer to a symmetric
double well. The spectra include localized
states for energies below the barrier energy $V_b=0.8$ eV and delocalized
states above the barrier.  For the specific parameters of
$V_0(x)$ considered, there are eight states with energies below the barrier. This number is
independent of the detuning voltage used. Furthermore, for asymmetric potentials,
all states with energies below the minimum energy of the higher well are
localized in the lower well, while states with energies above the minimum of the
higher well come in pairs and may not be fully localized. In the following, the lowest-lying state of this class is referred to as the metastable state with energy $E_m$.

%Sublofloat out

%\begin{figure}[hbt]
%	\captionsetup[subfigure]{labelformat=empty}
%	\subfloat[ \label{fig:potentials_ON} ]{\includegraphics[width=0.50\textwidth]{./pics/rrze/weak_potential_ON.pdf}} 
%	\hfill
%	\subfloat[ \label{fig:potentials_OFFMIX}]{\includegraphics[width=0.50\textwidth]{./pics/rrze/weak_potential_OFFMIX.pdf}} 
%	\hfill
%	\subfloat[ \label{fig:potentials_OFF}]{\includegraphics[width=0.50\textwidth]{./pics/rrze/weak_potential_OFF.pdf}} 
%	\hfill
%	\caption{Potentials $V_0(x)$, spectra and density of the eigenfunctions of the weakly bonded system for different detuning voltages $U_{d}$.}
%\label{fig:weak_potentials} 
%\end{figure}

\begin{figure}[htb] \begin{center} \includegraphics[width=0.5\textwidth]
		{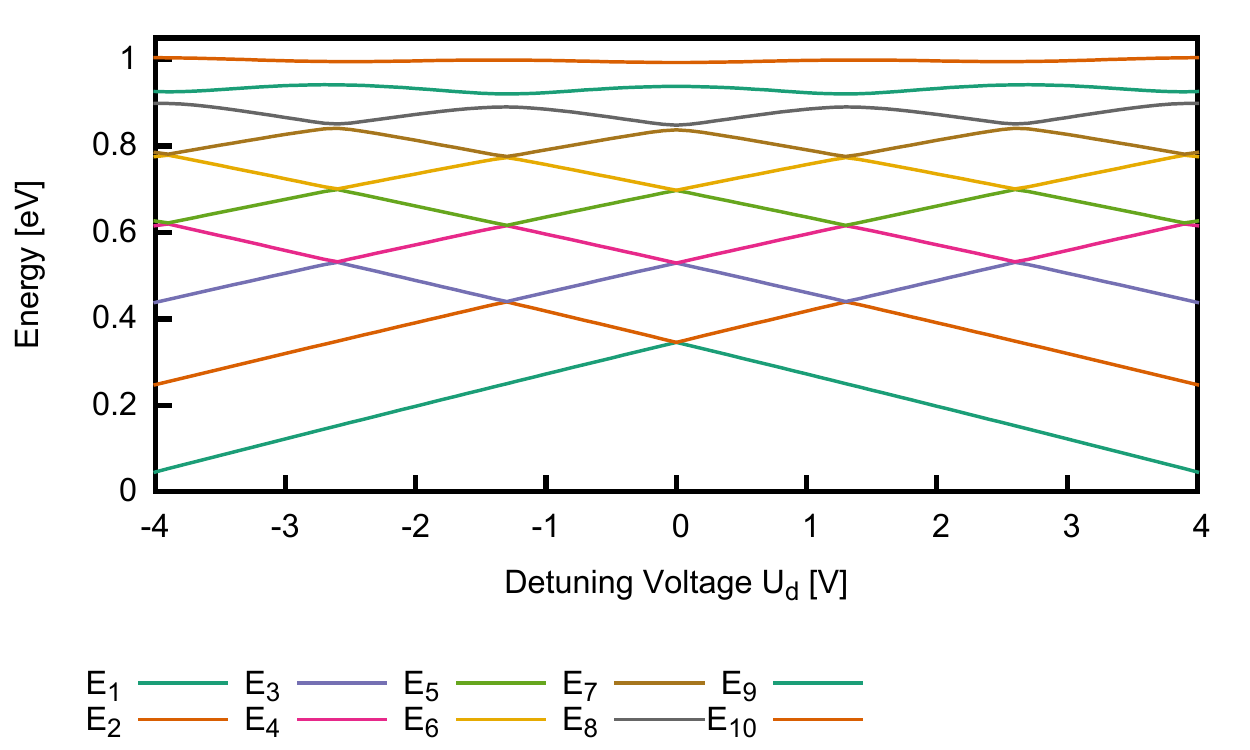} \caption{Eigenenergies of $h_0$ as a function of the detuning voltage.} 
			\label{fig:energydiagram_weak}.  \end{center}
	\end{figure}

The details of the spectrum, such as the energy spacing, depend on the detuning voltage.  In particular, the ground and the first excited
state are well separated in energy for the ON and OFF system, while they are rather close for the OFFMIX system. Upon changing the
detuning voltage, quasidegeneracies of the localized states occur, as illustrated in Fig.\ \ref{fig:energydiagram_weak}.  We refer to these
quasidegeneracies as resonances and denote the detuning voltage of the $n$-th resonance for positive (negative) detuning as $U_{n}^{+}$ ($U_{n}^{-}$). This classifies the OFF system as the first resonance with detuning voltage $U_d=U_{1}^{+}$ and the symmetric potential at the trigger point $U_{d}=0$ as the zeroth resonance state.

The diamond pattern of the energies in Fig.\ \ref{fig:energydiagram_weak} results from the oscillation of energies between adjacent states upon change of the detuning voltage. Specifically, a positive detuning ($U_{d} > 0$ V)  lifts the energies of the states localized in the left well
and lowers the energies of those localized in the right well.  At the resonance $U_{d} = U_{1}^{+}$ (corresponding to the OFF system), all states but
the ground state form tunneling pairs. Further detuning ($U_{d} > U_{1}^{+}$) moves the first excited state into the more stable well. In general,
the spectra of localized states for detuning values $U_{n}^{+} < U_{d} < U_{n+1}^{+}$ consist of $n+1$ states localized in the more stable well with energies lower than that of the metastable state ($E_m$) and $8 - (n+1)$ localized states with higher energies. 

\begin{figure}[htb] 
    \begin{center} \includegraphics[width=0.5\textwidth] {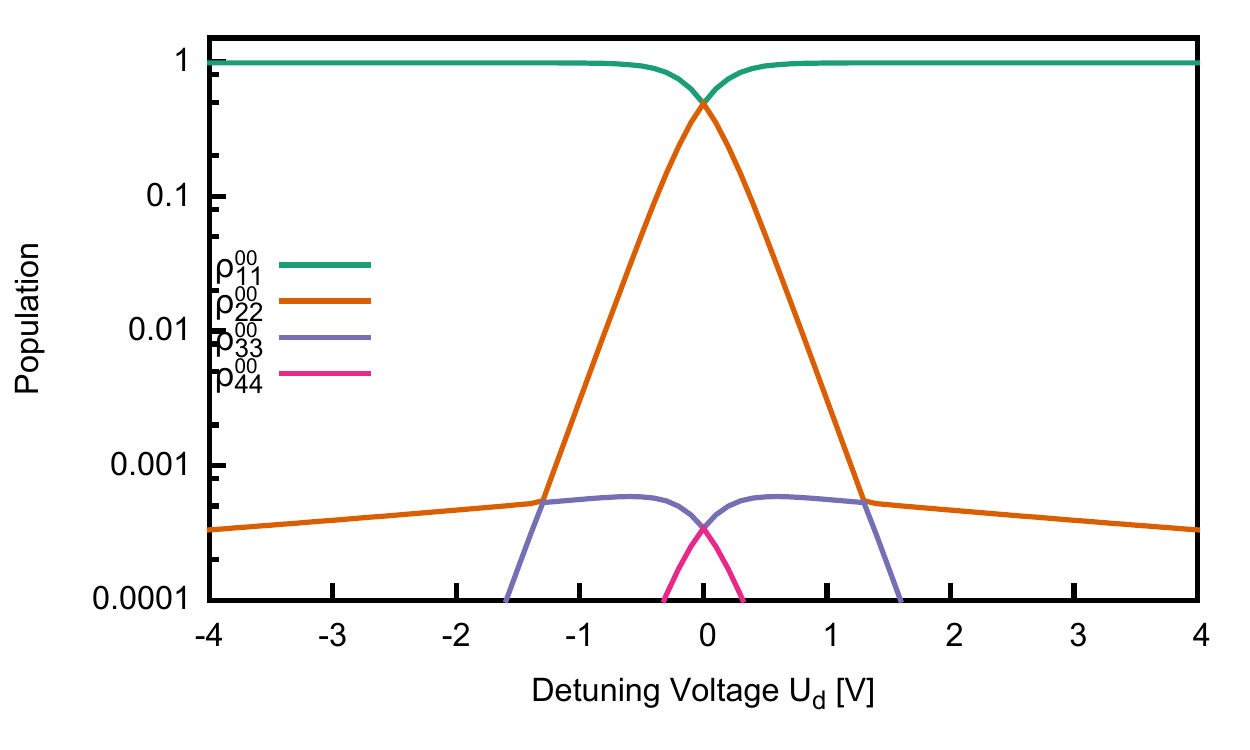} 
	\caption{Populations of the lowest-lying eigenstates, $\rho_{ii}^{00}$, for the molecular junction in equilibrium with the leads.}
    \label{fig:localizations} \end{center}
\end{figure}

Fig.\ \ref{fig:localizations} shows the population of the lowest-lying eigenstates,
$\rho_{ii}^{00}$ for the molecular junction in equilibrium, i.e.\ at zero bias voltage. At
the trigger point ($U_{d} = 0$ V) the two lowest-lying states of the tunneling pairs
are almost equally populated. For small detuning voltages, mostly the ground
and metastable states are populated. At the first resonance, the energy
difference between the metastable and the next higher-lying excited state minimizes so
that their population becomes very similar although it is significantly smaller
than that of the ground state.
% {\em warum brauchen wir hier eine neue Notation | Notation entfernt, in Figure ausgetauscht, in Figure ausgetauscht ???}

\subsubsection{Elementary charge transfer processes} \label{sec:transport-processes}

The basic mechanism of electron transport in the molecular junction model considered in this work involves a sequence of two
tunneling events. For positive bias, an electron with energy $\varepsilon_{k_{l}}$ in the left lead tunnels onto the molecule and occupies
the bridge state. Thereby, the proton is excited from the initial state $\ket{n,0}$ to $\ket{v,1}$ by energy exchange with the tunneling
electron. We denote the corresponding process by $P_{n,v}$.  In the subsequent event, the electron is transferred from the molecule to a
state in the right lead with energy $\varepsilon_{k_{r}}$, accompanied by another transition of the proton 

In the framework of Redfield theory, the energy is conserved within each individual process. The $P_{n,v}$ process defines a transport channel, which (for low temperatures $T \to 0$) is activated at threshold bias voltages $U_{nv} = \pm 2 \omega_{nv}$, where the sign accounts for excitations (+) and deexcitations ($-$). For finite temperatures the thresholds are reduced due to thermal broadening of the lead states. 

\begin{figure}[htb] \begin{center} \includegraphics[width=0.5\textwidth]{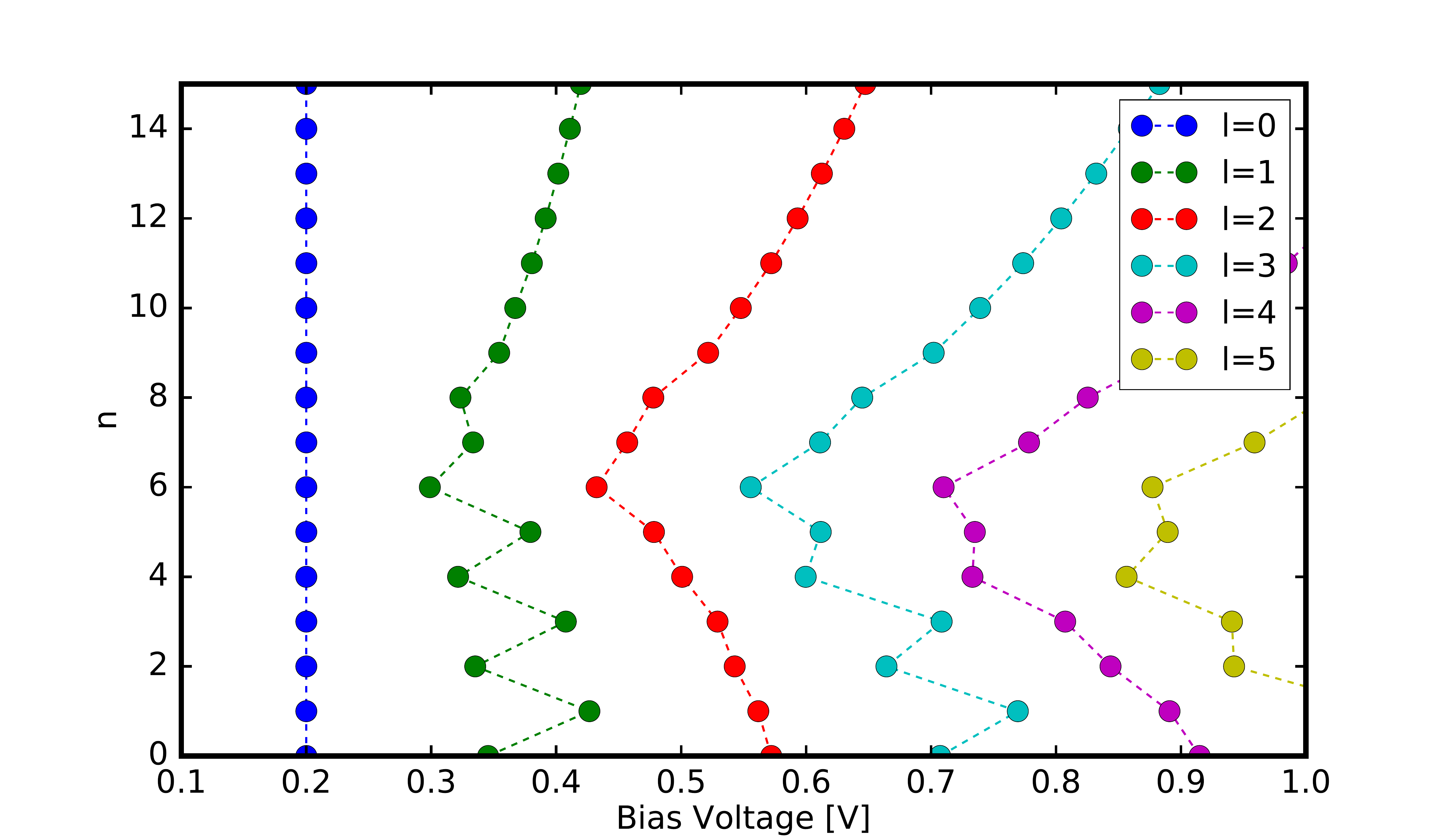}

\caption{$P_{n,n+l}$ processes for the OFFMIX systems represented as points over the respective threshold voltages. 
Processes of equal $l$ are depicted using the same color.}
\label{sub:transition_lines_none_zero} 
\end{center} 
\end{figure}

The threshold voltages $U_{nv}$ are closely related to the energy
spectra. This is illustrated for the OFFMIX system in Fig.\
\ref{sub:transition_lines_none_zero}, where the threshold voltages of the
processes $P_{n,n+l}=P_l$ are shown. The blue vertical line at $U_{on} = 2 \varepsilon_{0} = 0.2$ V
corresponds to the onset, where all elastic processes $P_{n,n}$ are enabled
simultaneously. All other lines correspond to inelastic processes with $l>0$.

In a double well potential, transitions between localized states can be
classified as intra- and inter-well transitions, which correspond in the OFFMIX
system to processes with odd (e.g.\ $P_{n,n+1}=P_1$) or even (e.g.\ $P_{n,n+2}=P_2$) value of $l$, respectively.  Fig.\
\ref{sub:transition_lines_none_zero} demonstrates that the threshold voltages
for the two types of processes follow different patterns.  Inter-well processes
$P_{1}$ exhibit for small $n$ a zig-zag pattern, which is a result of the state
pairing in the double well.  Thereby, small threshold voltages correspond to
transitions among pairs.  On the other hand, the linear pattern for intra-well
processes $P_{2}$ for small $n$  is related to the decreasing energy gap
between adjacent states localized in the same well. In particular, the energy differences are minimal around the barrier of the double-well potential. As a result, transport processes involving states close to
the barrier are activated prior to those that involve states localized deeper in the potential well.

The analysis above has only considered the opening of the different transport channels. The
extent to which open channels actually contribute to charge transport is
determined by the nonequilibrium population of the states and the transition
rates (cf.\ Eq.\ (\ref{eq:explicit-redfield-energy})). The latter involve, besides the molecule-lead coupling, the transition elements $S_{m,v} =
\left| \bra{m,0} s(x) \ket{v,1} \right|^{2}$ with the switch function $s(x)$. The transition elements
take the form 
\begin{equation} S_{m,v} = \left|
	\frac{1+\sigma}{2} \delta_{m,v} - \frac{1 - \sigma}{2} \int dx
	\psi_{v}^{1}(x) \psi_{m}^{0}(x) \tanh(x) \right|^{2}, \label{eq:frank-condon}
\end{equation} 
where $\psi_{k}^{i}(x)= \Braket{x,i|\psi_{k} }$. 

For low bias voltages, transport is dominated by elastic 
processes $P_{0}$ involving localized states. For states strongly localized  in the wells, the switch function $s(x)$ is nearly constant. In an asymmetric double-well potential, i.e.\ for $U_{d} \neq 0$, the transition
elements for elastic processes are obtained from Eq.\ (\ref{eq:frank-condon})
as $S_{m,m}^{L} \approx 1$ for the left and $S_{m,m}^{R} \approx \sigma^{2}$
for the right well, respectively.  Thus, the rates of elastic processes in the right well
are downscaled by a factor of $\sigma^{2}$.  On the other hand, for delocalized states, which are important for higher bias voltages,
the wave functions can be assumed to be approximately constant, yielding for elastic transitions $S_{m,m}^{D} \approx \left( 1 + \sigma \right)^{2} \mathbin{/} 4$.

\begin{figure}[htb] \begin{center} \includegraphics[width=0.5\textwidth]
		{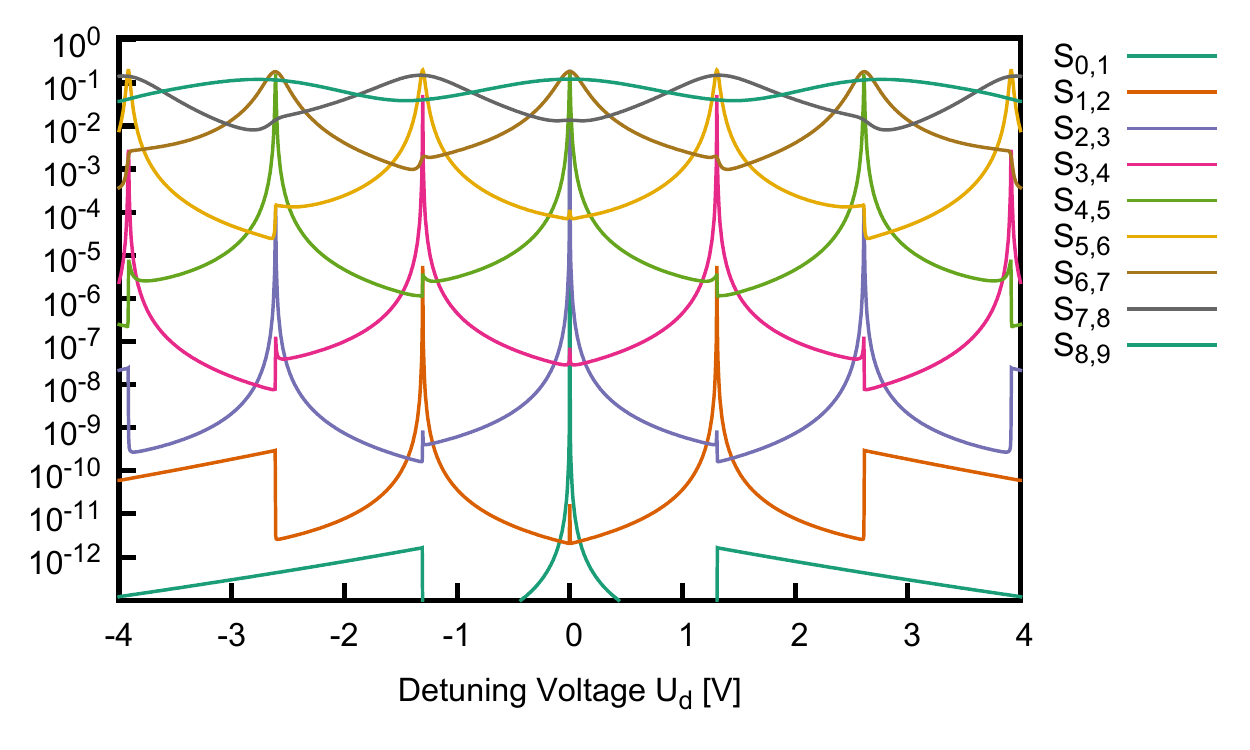} \caption{Transition
			elements  $S_{n,n+1}$ of processes $P_{n,n+1}$ as a function of the detuning voltage} 
			\label{fig:proton_electron_coupling}
		\end{center} \end{figure}

Fig.\ \ref{fig:proton_electron_coupling} shows the transition elements for
the inelastic processes $P_{1}$ as a function of the detuning voltage.
The rates of these processes are rather small for lower-lying states due to the fact that these states localize in different potential wells.
For higher-lying states, which become relevant at high energies/voltages, 
inelastic processes play a more significant role. The reason for this is that the wavefunctions of these states, which are mainly localized in one of the potential wells, exhibit also small tails in the other potential well. The contribution of these tails to the integral in Eq.\ (\ref{eq:frank-condon}) increases with higher energy, therefore enhancing the role of these processes. Furthermore, the transition elements of paired states can also give pronounced contributions resulting in the peaks depicted in Fig.\ \ref{fig:proton_electron_coupling}, which occur at quasidegenaracies of the energies (see Fig.\ \ref{fig:energydiagram_weak}).

\subsubsection{Transport properties}
\label{sec:biasing_weak}

\begin{figure}[htb] 
    \begin{center} 
        \includegraphics[width=0.5\textwidth] {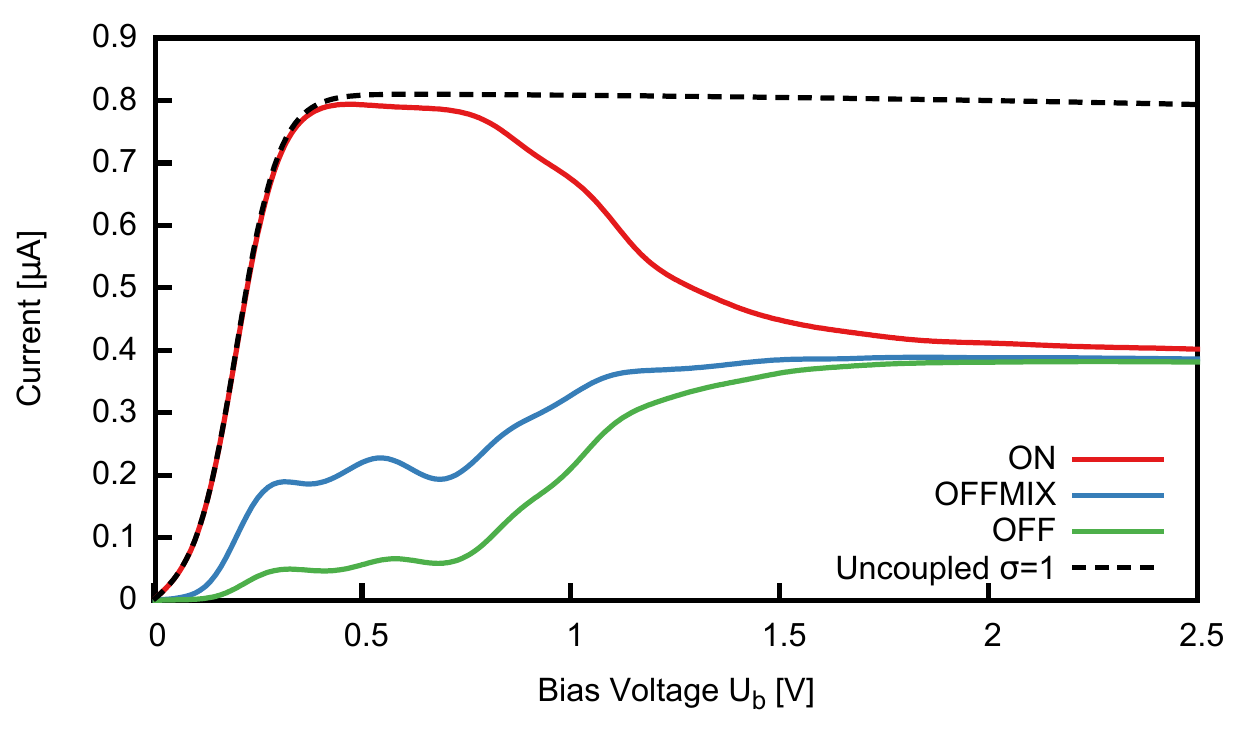}
      \caption{Current voltage characteristics of the ON, OFFMIX, OFF, and uncoupled ($\sigma=1$) systems. }
      \label{fig:current-voltgate-weak}
    \end{center}
\end{figure}

The current-voltage characteristics of the ON, OFFMIX, and OFF systems are shown
in Fig.\ \ref{fig:current-voltgate-weak}. For comparison, the current for the
system without coupling between the electronic and proton degrees of freedom
(in the following referred to as the uncoupled system) is also depicted. The currents
of the coupled systems exhibit pronounced differences for small bias voltages but
tend to the same saturation value of $I_{\rm sat} = 0.4$ $\mu A$ for bias
voltages $U_{b} > U_{\rm sat} = 1.5$ V, a value significantly smaller than that found for the uncoupled system. All coupled systems exhibit structures at low bias voltages, which are more pronounced for the OFF and OFFMIX systems.
However, while in these systems the saturation current is approached from below, the ON
system shows a more complex behavior. Specifically, this system exhibits a strong increase of the current at low bias voltages closely resembling that of the uncoupled system. This is followed by a plateau and a pronounced negative
differential conductance (NDC) towards the saturation limit for larger bias voltages. In addition, the current of the ON system is always larger than that of the other two coupled
systems.

\begin{figure}[htb] 
    \begin{center} 
        \includegraphics[width=0.5\textwidth]{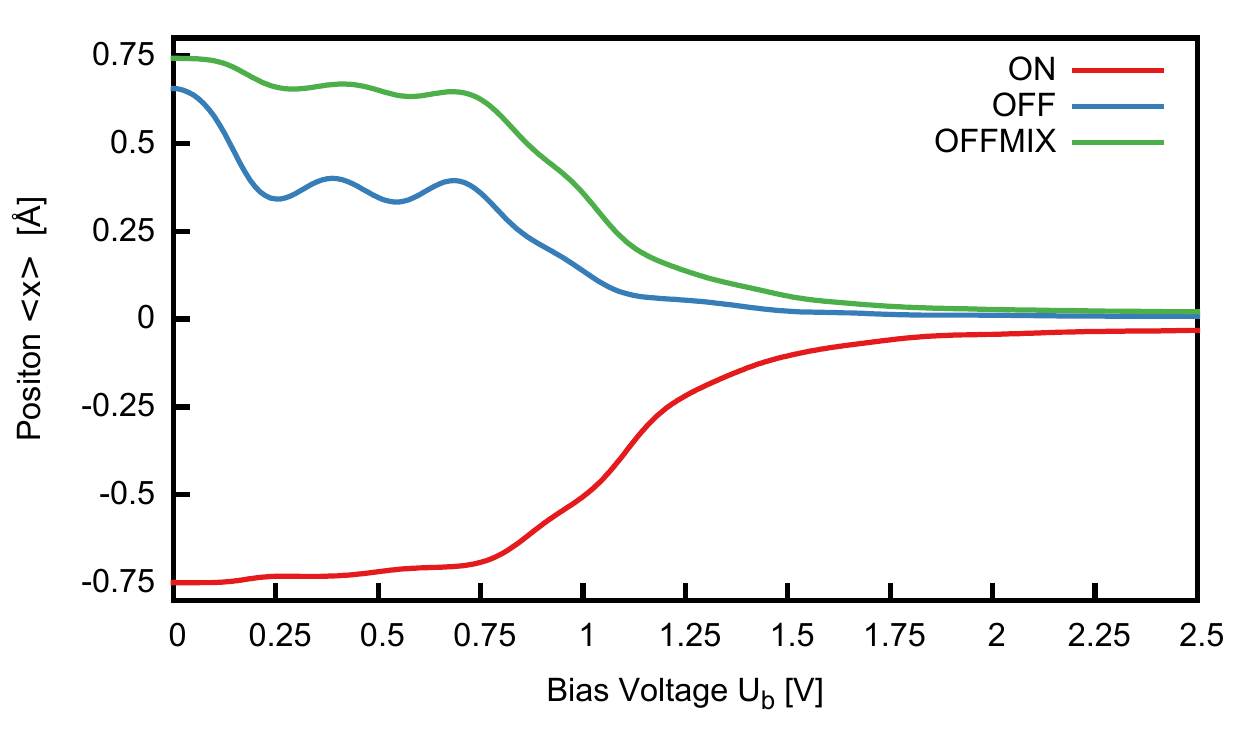} 
        \caption{Average position of the proton $\Braket{x}$ as a function of the bias voltage for the ON, OFF, and OFFMIX systems. }
            %The additional curve labeled with ``minimal'' corresponds
            %to the density matrix of the OFFMIX system, which is computed by the minimal set of $M=8$ energy eigenstates and
            %resembles the full calculation with $M=60$ for low voltages very well. 

        \label{fig:position_characteristics} 
    \end{center}
\end{figure}

\begin{figure}[htb] \begin{center}
		\includegraphics[width=0.5\textwidth]{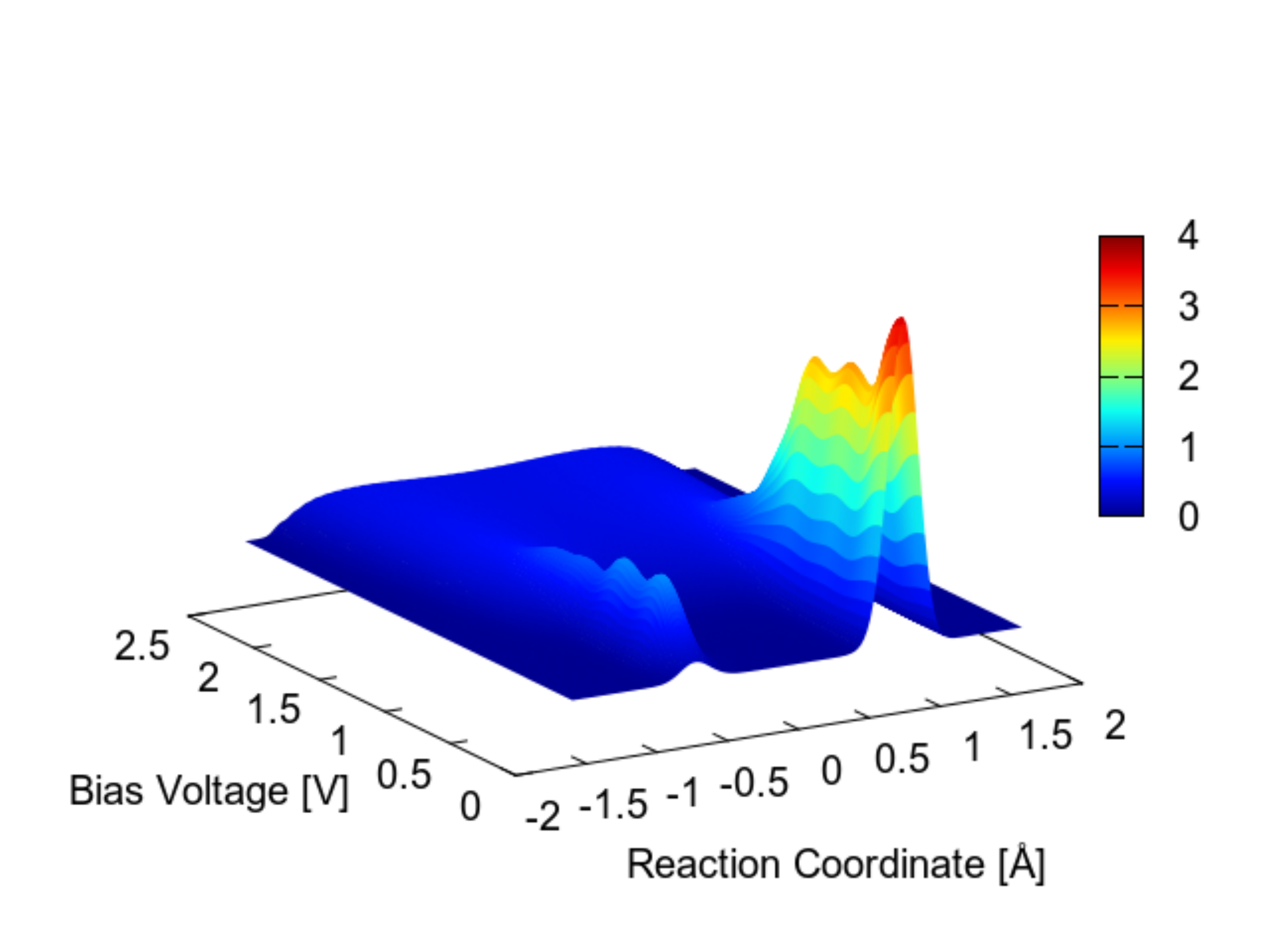}
		\caption{Probability distribution of the proton position $\rho(x)$ as a function of the bias voltage for the OFFMIX system.}
%			For low voltages, vibrational states in the shallow well %(left side) are excited increasing 
%			its population. The proton oscillates slightly between the %wells due to
%			stimulated tunneling processes in the intermediate region. %After			exceeding the saturation voltage $U_{sat}$, the proton %is uniformly
%		distributed over the delocalized area.} 
		\label{fig:weak_rhox}
	\end{center} \end{figure}

The behavior of the current-voltage characteristics can be rationalized on the basis of the position of the proton.  Fig.\ \ref{fig:position_characteristics} depicts the average position of the
proton ($\Braket{x}$) for the ON, OFFMIX, and OFF systems and Fig.\ \ref{fig:weak_rhox} shows the probability distribution of the proton ($\rho(x)$) for the
case of the OFFMIX system. At a given bias voltage $U_{b}$, the position
of the proton determines the effective molecular-lead coupling via the
switch function $s(x)$, cf.\ Eq.\ (\ref{eq:switch_function}). As shown in Fig.~\ref{fig:position_characteristics} the average position of the proton for the three coupled systems resembles the overall behavior of the current-voltage
characteristics. Specifically, an increase of the bias voltage drives the average position of the proton
from the equilibrium position in one of the wells at $U_{b}=0$ to
the center of the double well potential $x_{\rm sat}=0$ {\AA} at $U_{b}>1.5$ V, indicating that it is fully delocalized.

A more detailed inspection of the average position of the proton as a function of the bias voltage shows that for low bias voltages, the proton remains
localized at the global minimum in the ON and OFF systems, with the elastic transition $P_{1,1}$ of the ground state of the proton being the dominant contribution to the transport process. Since the transition element for the localization in the left well is $S_{1,1}^{L}\approx 1$, the effective molecule-lead coupling of the ON system corresponds to that of the uncoupled case ($\sigma=1$). This explains the similarities found in the conductance of both systems in the range of bias voltages $U_{b}<0.5$ V. In contrast, due to the localization of the proton in the right well, the molecule-lead coupling in the OFF system is downscaled by a factor of
$S_{1,1}^{R} \mathbin{/} S_{1,1}^{L} \approx \sigma^{2} \approx 0.01$, which
agrees with the observed ratio of the currents in the ON/OFF systems.

In the case of the OFFMIX system, in addition to the ground state the metastable
state is also significantly populated at $U_{b}=0$ V as discussed in Sec.\
\ref{sec:weak_system_analysis}, which shifts the
equilibrium position of the proton slightly towards $x_{\rm sat}$.  As a result, the current at low
bias voltages in this system involves contributions of the ground state process
$P_{1,1}$ and the metastable state process $P_{2,2}$ resulting in values of
the current that are between those of the ON and OFF systems. 

At bias voltages larger than the onset  value of $U_{\rm on}=0.2$ V, excited
vibrational states are populated. In the ON system, the first vibrational excited state is
located in the same well as the ground state.  As a result, the excitation of
this state hardly changes the position of the proton and the molecule-lead
coupling and the current remains at a plateau.
In contrast, this situation is reversed in the OFFMIX case and, as a result, the position of the proton changes notably in this system, as shown by the redistribution of population among the two wells in Fig.\ \ref{fig:weak_rhox}. 

For  higher bias voltages $U_{b} > U_{\rm on}$, inelastic transport processes
are activated. In the OFFMIX system, these processes result in pronounced
oscillations of the average position and the current as a function of the bias
voltage. As discussed in Sec.\ \ref{sec:transport-processes}, these
oscillations are related to contributions of the inelastic processes $P_{n,n+1}$
and $P_{n,n+2}$.  
The ON and OFF systems also exhibit oscillatory structures in this bias voltage
range, but these are less pronounced due to the more localized character of the
states populated.

%{\em Beschriftung der Achsen ueberarbeiten!}

Increasing the bias voltage further towards the saturation value $U_{\rm sat} \approx 1.5$ V, activates inelastic processes that excite the proton
above the barrier. In particular, transitions from localized to delocalized states close to the potential barrier become
important. As a result, the proton is distributed almost uniformly over the translocation path as shown in Fig.\ \ref{fig:weak_rhox} for the
OFFMIX system (a similar result is found for the ON and OFF systems) leading to an average position of $ \Braket {x}=0$ {\AA}. Since in this voltage range the
accesible vibrational states are populated almost equally, the effective molecular-lead coupling is given by the average of the switching
function, $ \int_{-\infty}^{\infty} s(x) dx =  \frac{1 + \sigma}{2}  \approx 0.5$. This result explains the reduction of the current of the ON system from the plateau value to the saturation value corresponding to a pronounced NDC behavior.
Based on this reasoning, any system with a detuning voltage of $U_d < 0$ V is expected to show this qualitative behavior.

The results discussed so far demonstrate that employing a gate voltage different conductance states of the molecular bridge in the junction can be achieved exhibiting low current (OFF system) or high current (ON system). This is, however, restricted to a certain range of bias voltages. For high bias voltages, the ability to control the conductance state of the molecular bridge in the junction using the gate potential is lost because delocalized states above the barrier are populated.  However, the control of the conductance state of the molecular bridge in the junction can be recovered if the relaxation of the proton motion due to interaction with other vibrational modes of the molecule, phonons in the electrodes or solvent is taken into account.

\begin{figure}[htb] 
\begin{center} 
\includegraphics[width=0.5\textwidth] {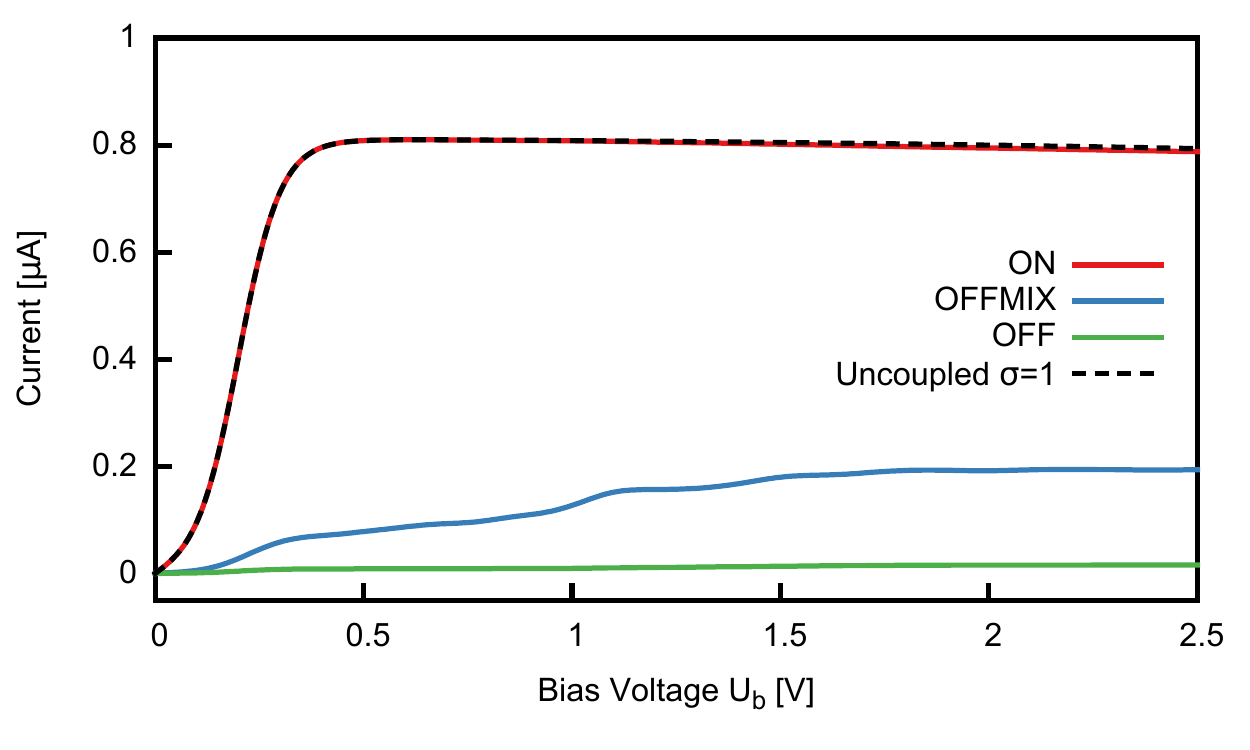} 
\caption{Current-voltage characteristics of the ON, OFFMIX, and OFF systems calculated including the coupling 
to a harmonic bath ($\eta=0.001$). For comparison, the current of the system with no coupling between the electronic and proton degrees of freedom ($\sigma=1)$ is also shown. }
      \label{fig:position_characteristics_env}
\end{center} 
\end{figure}

Fig.\ \ref{fig:position_characteristics_env} shows the current-voltage characteristics of the ON, OFF, and OFFMIX systems including relaxation of the proton motion. This effect was modeled by coupling the system to a harmonic bath, Eq.\ (\ref{eq:harmonic_bath}) with coupling
strength $\eta=0.001$ corresponding to weak coupling. For comparison, the current of the system without coupling between the electronic and proton degrees of freedom ($\sigma=1)$ is also depicted.

\begin{figure}[htb] \begin{center} \includegraphics[width=0.5\textwidth]{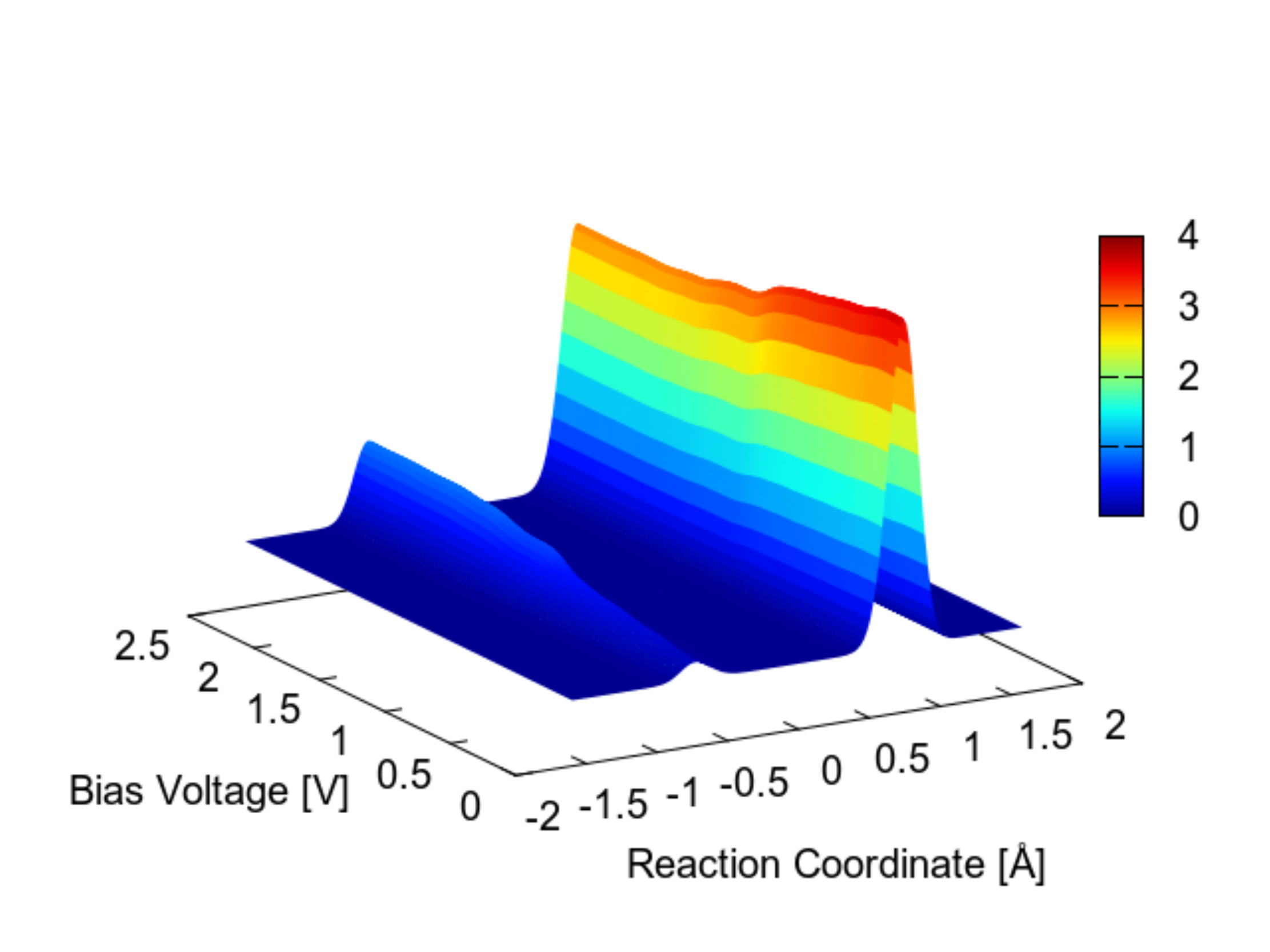} \caption{Probability distribution of the proton position
			$\rho(x)$ of the OFFMIX system calculated for a system-bath coupling of $\eta=0.001$.}
%			The increase of the bias voltage populates slight %monotonously the left 
%			upper well}
			\label{fig:rhox_env} \end{center} \end{figure}

The results in Fig.\ \ref{fig:position_characteristics_env} show that for the three systems the current is different over the whole range of bias voltages. This result, which differs from that obtained in the absence of a system-bath coupling (see Fig.\ \ref{fig:current-voltgate-weak}), is
a consequence of the relaxation of the proton motion, which results in the population of low-lying, localized states even for large voltages. This is demonstrated in Fig.\ \ref{fig:rhox_env} for the probability distribution of the proton position  $\rho(x)$ of the OFFMIX system. A detailed analysis of the
transition elements $X_{k,k^{\prime}}$ of the harmonic bath reveals that the relaxation is most effective among delocalized states and states localized in the same well below the barrier.

\subsubsection{Functionality} 
\label{ref:weak_functionality} 

\begin{figure}[htb] 
    \begin{center} \includegraphics[width=0.5\textwidth] {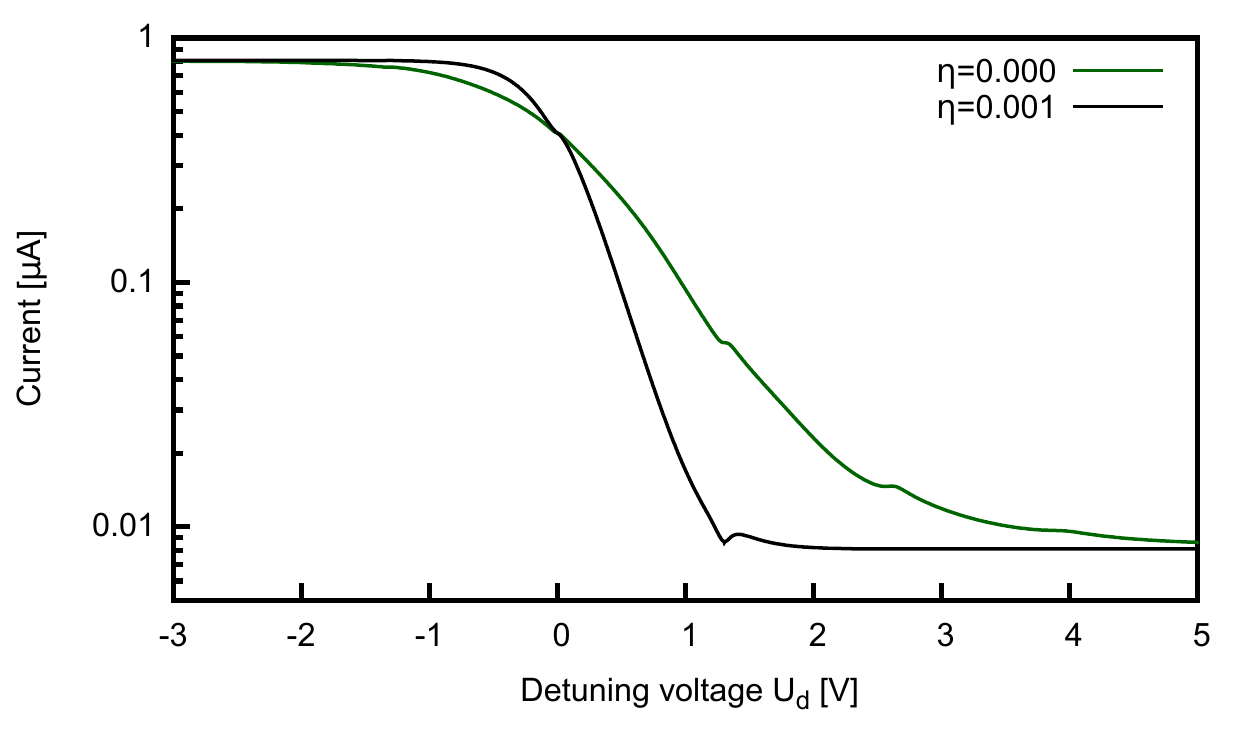} 
	    \caption{Current as a function of the detuning voltage $U_d$ for a constant bias voltage $U_{b}=0.7$ V with and without system-bath coupling ($\eta$). At a detuning voltage $U_d \sim 1-2$ V the ``ON'' state switches to the ``OFF'' state.}
        \label{fig:switching_weak} 
    \end{center}
\end{figure}

The dependence of the current on the gate voltage can be utilized to realize a
molecular transistor. This is demonstrated in Fig.\ \ref{fig:switching_weak},
which shows the current as a function of the detuning voltage for a fixed bias
voltage $U_{b} = 0.7$ V. This value ensures that even without relaxation
(i.e.\ $\eta =0$), the proton is localized in one of  the wells.  The results
show a transition from a high current (``ON'') to a low current (``OFF'') state at a detuning
voltage $U_d \sim 1-2$ V indicating the translocation of the proton. This
transition voltage is slightly off the expected value of $U_d=0$. This is
due to the different contribution of elastic processes in the left and right wells.
Although the population of the uplifted left well decreases with the increase of
the detuning voltage $U_{d}>0$ V, the corresponding weaker transport processes
in this well have a pronounced impact on the current since their contribution
is enhanced by a factor of $1/\sigma^{2} = 100$. In the reverse situation for $U_{d} < 0$ V, the transport processes
in the uplifted right well are negligible owing to this factor.
Therefore the
current saturates asymmetrically with respect to the trigger point.

Coupling to the harmonic bath results in a shrinking of the transition zone because the bath
suppresses the population of the metastable state as it facilitates the relaxation of the proton
to the ground state. This results in a defined conductance state already at
small detuning voltages. The current also exhibits structures at certain detuning voltages (see Fig.\
\ref{fig:switching_weak}) which, as discussed in Sec.\ \ref{sec:transport-processes}, correspond to the $U_{n}$ resonances. 

So far, we have neglected the effect of the bias voltage on
the proton potential, corresponding to a value of $\phi=\pi/2$ (cf.\ Eq.\
\ref{eq:external_field}). In the following, we study the more general case and
show that the influence of the bias voltage on the proton potential results in
a diode-like behavior of the molecular junction.  For a general angle $\phi$
and a given gate voltage $U_g$, the bias voltage at which the potential
for the intramolecular proton transfer is a symmetric double well (the so-called trigger point) is given by
\begin{equation} U_{\rm b, sym} \left( U_{g} \right) = \frac{L_{m}
		\varepsilon_{d}}{2 l \cos(\varphi)} - \frac{L_{m}}{d} U_{g} \tan( \varphi ).
		\label{eq:symmtric-voltage-bias} 
\end{equation} 
Therefore, the global minimum of the potential is located in the left well for $U_{b} < U_{\rm b,sym}$ and otherwise in the right well. 

\begin{figure}[htb] \begin{center} \includegraphics[width=0.5\textwidth]
		{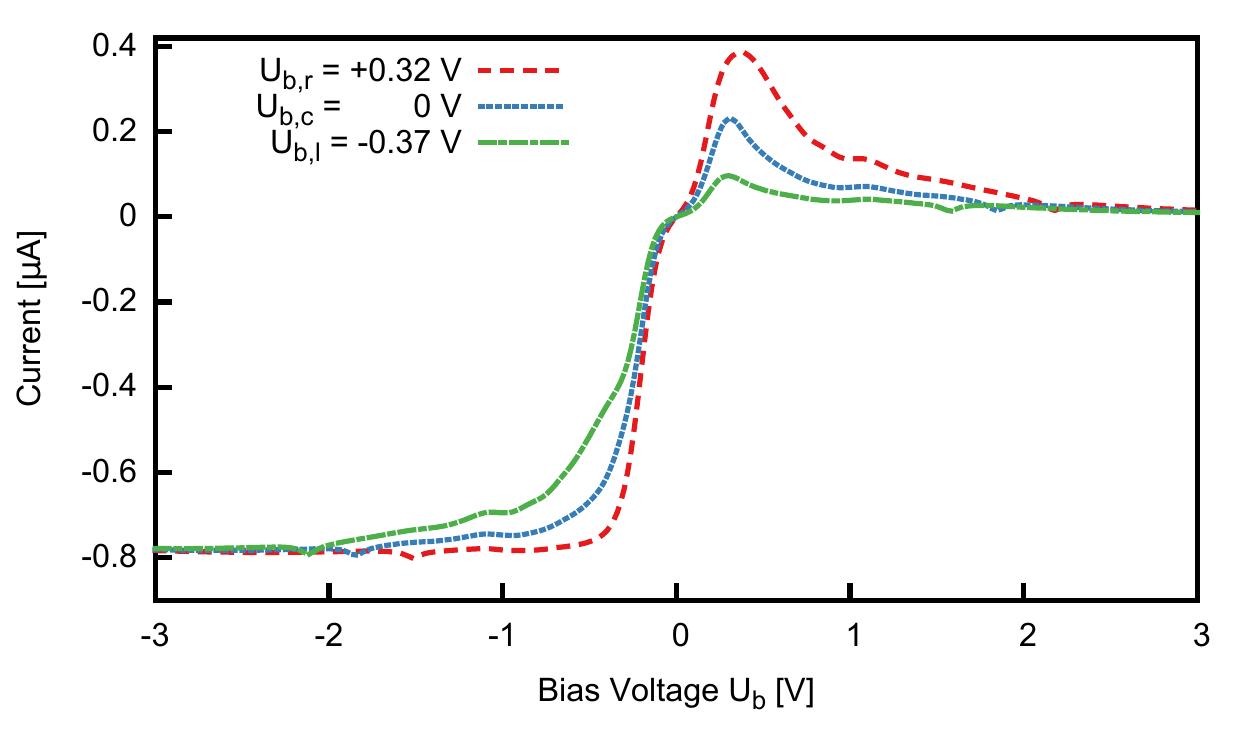} \caption{Current-voltage characteristics for different values of the gate potential (see text) corresponding to different trigger points and for a system bath coupling $\eta=0.001$ calculated for an angle $\phi = \pi \mathbin{/} 4$ between the proton translocation path and the bias field. } 
\label{fig:weak_diode} \end{center} 
\end{figure}

Fig.\ \ref{fig:weak_diode} shows the current-voltage characteristics for an angle of $\phi = \pi \mathbin{/} 4$, system-bath coupling $\eta =0.001 $ and for different values of the gate potential corresponding to different trigger points, 
$U_{b, l} \left(3.2 \right) = -0.37$ V, $U_{b, r}\left( 2.5 \right) = 0.32$ V and $U_{b,c} \left(2.82 \right) = 0$ V. 
The current-voltage characteristics shows the typical behavior of a diode with a significant current for negative bias voltages and vanishing current for large positive bias voltages. The details depend on the chosen gate voltage and, as a consequence, on the trigger point. 

As shown in Fig.\ \ref{fig:weak_diode}, all systems considered exhibit a NDC effect for positive bias voltages, which is most pronounced for the system with the trigger point located at $U_{b} = U_{b, r}$.  This system corresponds to an ``ON''
conductance state for the bias voltage $U_{b} \le U_{b,r}$ resulting in a
high current at the onset of the positive bias. For larger bias voltages $U_{b} \ge
U_{b,l}$ the current decreases since the ``OFF''
state stabilizes and becomes increasingly populated. Although the effect is similar to that found for the ON system discussed in Sec.\ref{sec:biasing_weak} the cause of the NDC in this case does not relate to the reduction of the effective molecular-lead coupling as a consequence of the
delocalization of the proton but rather to the change of the global minima.  
The rise of the current at positive voltages is less pronounced in the system with the trigger point at $U_{b,l}$.  In this system, the conductance state is
already changed to ``OFF'' at $U_{b}=U_{b,l}$  before the bias voltage becomes positive, which results in a smaller current. 
In conclusion, the position of the trigger point $U_{\rm b,sym}$ determines the increase in the current at positive bias voltages due to contributions of the transport processes in the left well, which cause a significant enhancement of the current. For larger bias voltages the contribution of these processes decreases as a consequence of the destabilization of the left potential well.

\subsection{Stronger hydrogen bonded, low-barrier systems  \label{sec:comparison_all_systems}}
In this section, we expand our analysis and consider systems with smaller barriers for the intramolecular proton transfer reaction and stronger hydrogen bonding. Specifically, we consider two representative systems with the potentials shown in Fig.\
\ref{fig:smaller_potentials}. The parameters are given in table \ref{tab:parameters}. For simplicity, in this section we neglect the
influence of the bias voltage on the potential $V_0(x)$ of the proton motion. This corresponds to a situation, where the direction of the
translocation path is parallel to the gate field (cf.\ Fig.\ \ref{fig:angle}), i.e.\ $\phi = \pi \mathbin{/} 2$.

\subsubsection{Equilibrium properties}

\begin{figure}[hbt]
	\includegraphics[width=0.45\textwidth]{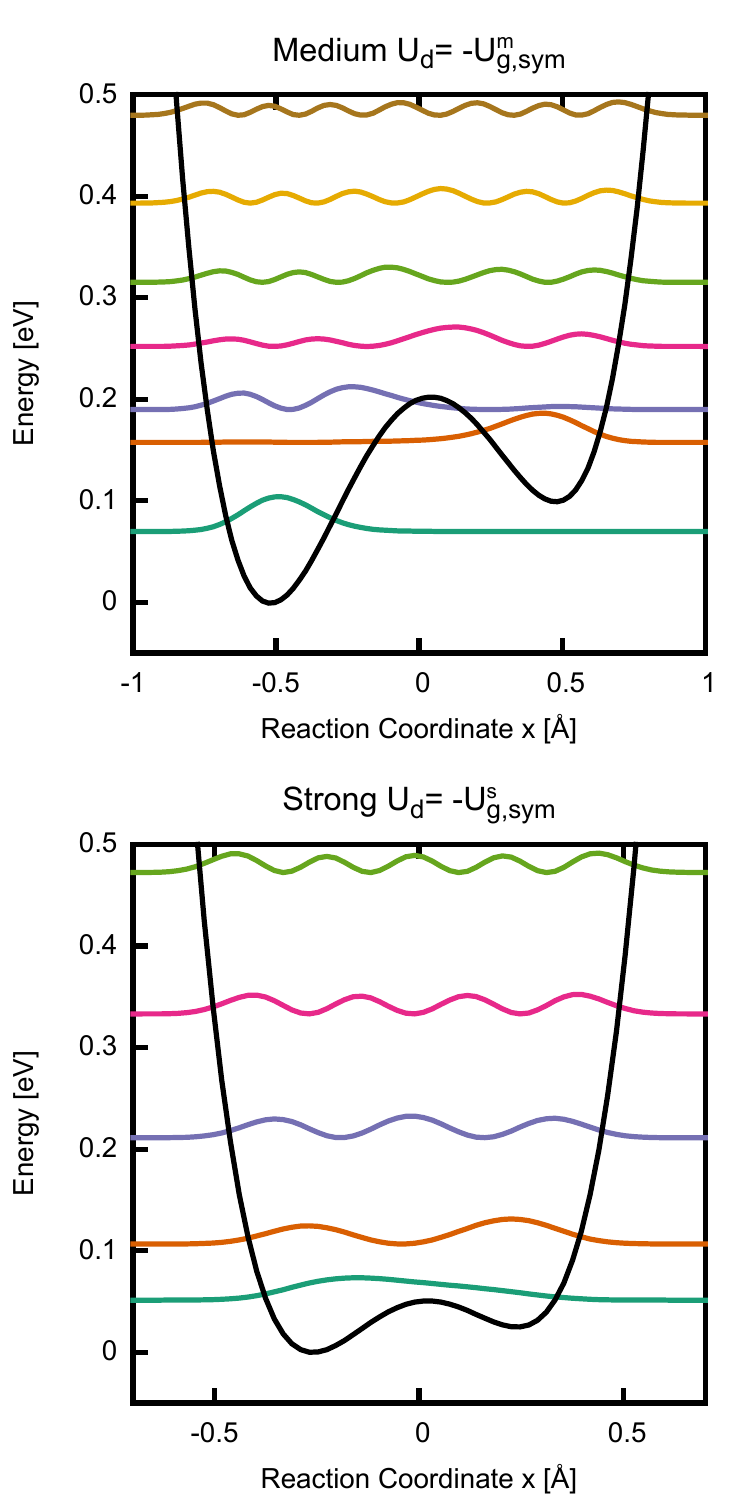}
	\caption{Potentials $V_0(x)$, spectra and density of the eigenfunctions of the medium and strongly hydrogen bonded systems.}
\label{fig:smaller_potentials} 
\end{figure}

The energy spectra and character of the eigenstates of the two strongly bonded
systems shown in  Fig.\ \ref{fig:smaller_potentials}  differ
significantly
from those of the weakly bonded system considered in Sec.\ \ref{sec:weak_system_analysis}. Specifically, in the medium
bonded system, only two localized states exist with energies below the barrier.
In the strongly bonded system already the ground state has an energy above the
very small barrier and thus no localized eigenstates exist. The trigger points
of the medium and the strongly bonded system are $U_{\rm g,sym}^{m} = 1.0$ V and
$U_{\rm g,sym}^{s} = 0.5$ V. 

%\begin{figure}[hbt]
%	\captionsetup[subfigure]{labelformat=empty}
%	\subfloat[\label{fig:potentials_medium} ]{\includegraphics[width=0.50\textwidth]{./pics/rrze/compare/comparison_potentials_medium.pdf}} 
%	\hfill
%	\subfloat[\label{fig:potentials_small}]{\includegraphics[width=0.50\textwidth]{./pics/rrze/compare/comparison_potentials_small.pdf}} 
%	\hfill
%	\caption{Potentials $V_0(x)$, spectra and density of the eigenfunctions of the medium and strongly hydrogen bonded systems.}
%\label{fig:smaller_potentials} 
%\end{figure}

\begin{figure}[htb] 
\begin{center} \includegraphics[width=0.5\textwidth]
		{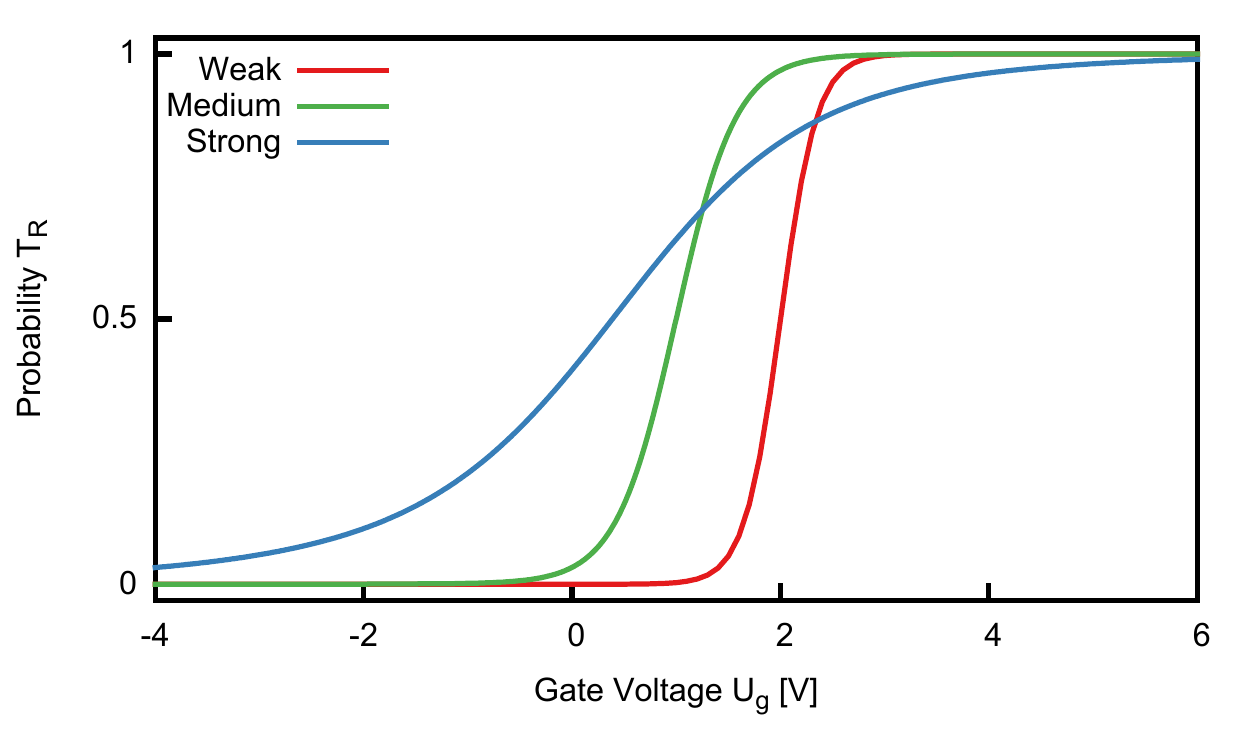} \caption{Probability of finding the proton in the right half-space ($T_{R}$) as a function of the gate voltage for the three systems corresponding to weak, medium, and strong hydrogen bonding.}  
\label{fig:tunneling_potentials} 
\end{center} 
\end{figure}

As a consequence, the localization of the proton in the equilibrium state is also quite different in the three systems, as indicated by the probability of finding the proton in the right half-space, 
$T_{R} = \int_{0}^{\infty} \rho(x) dx$, depicted in Fig.\  \ref{fig:tunneling_potentials} as a function of
the gate voltage $U_g$.  At their respective trigger points, the proton is
equally distributed, $T_{R}=0.5$, owing to the symmetry of the potentials.
While the weakly and medium bonded system exhibit a rather sharp transition
between wells, in the strongly bonded
system it is spread over a large range of gate voltages. This is
due to the fact that this system has no
localized states at the trigger point and thus a relatively large gate voltage is required to
achieve localization in one of the wells. 

\subsubsection{Transport properties}

\begin{figure*}[htb]
\begin{center} 
\includegraphics[width=1.0\textwidth] {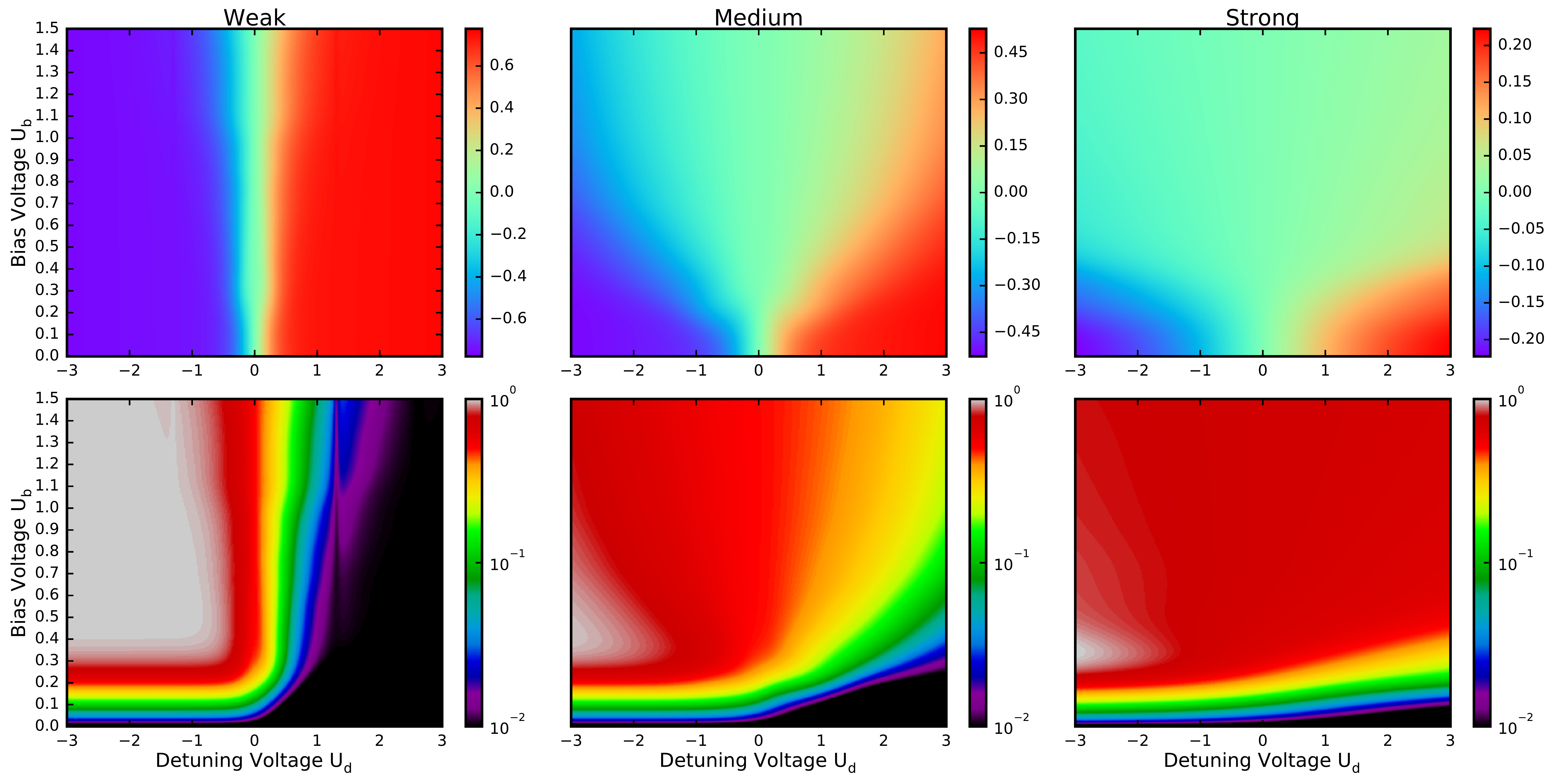} 
		 \caption{Average position $\langle x \rangle$ (top) and corresponding current ratio $I\mathbin{/}I_{max}$ (bottom) as a function of detuning and bias voltage for the three systems corresponding to weak, medium, and strong hydrogen bonding. The results have been obtained including a system-bath coupling with strength $\eta=0.001$.} 
\label{fig:heatmap_no_env} 
\end{center} 
\end{figure*}

Fig.\ \ref{fig:heatmap_no_env} shows the average
proton position $\langle x \rangle$ (upper panels) and the current (lower panels) as a function of the
detuning and bias voltage for the weakly, medium, and strongly bonded systems. %in
%comparison to those of the weakly bounded system already discussed in Sec.\ \ref{sec:weak_system_analysis}. 
The results have been obtained taking relaxation of the proton into account ($\eta = 0.001$). For better comparability, the current is shown relative to its maximal value, which is given by $I_{max}^{w}=I_{max}^{m} = 0.81$ $\mu A$ 
and $I_{max}^{s} = 0.59$ $\mu A$ for the weakly, medium, and strongly bonded system, respectively.

The results for the average proton position show clearly the regions of localization in the left (blue) and the right (red) well. These regions are separated by the funnel-shaped region of delocalization (green) centered at the trigger point. The shape of this region is quite different for
the three systems investigated. Specifically, in the weakly bonded system the proton is localized for most detuning and bias voltages and the delocalization region extends over a narrow range of small detuning voltages corresponding to a symmetric double-well potential. In the more strongly bonded systems, on the other hand, the proton is always delocalized for sufficiently high bias voltages. 

As a result of the distinct localization pattern exhibited by the proton for the different systems, their current-voltage characteristics differ
significantly. While the weakly bonded system shows a distinct dependence of the current on the detuning voltage, this is less
pronounced in the medium bonded system and almost negligible in the strongly bonded system.  As a consequence, only the weakly bonded system,
which exhibits a significant barrier, can fulfill the functionalities of a transistor or a diode.

\section{Conclusions}\label{sec:conclusion}

In this paper we have studied the influence of an intramolecular  proton transfer reaction on the conductance of a molecular junction. To this end, we have used a generic model for a proton transfer reaction with parameters motivated by our previous first-principles studies and employed a quantum master equation approach at the level of Redfield theory to solve the transport problem.  

The results show that it is possible to control the conductance state of molecular junctions using a proton transfer reaction in combination with external electric fields. Depending on the location of the proton, the junction exhibits high or low current. Considering different parameter regimes, which range from weak to strong hydrogen bonding and include situations with high or low barriers separating the reactant and product of the reaction, we have identified necessary preconditions for achieving control. We have also demonstrated that the proton transfer mechanism can be utilized to achieve functionality. Employing external fields of a gate or the lead electrodes, the current-voltage characteristics of the molecular junction in systems with a weak hydrogen bond and a significant energy barrier for the proton transfer resemble those of a transistor or a molecular diode. 

In the present paper, we have used a generic model and focused on the steady state transport properties. Another interesting question concerns the time-dependent transport properties, e.g.\ how the steady state of the molecular junction is reached. For a realization of a molecular switch, for example, an intriguing question is how the low conductance state transforms into the high conductance state and vice versa upon change of the external electric field. This question, as well as an extension of the model towards a first-principles based description will be the topic of future investigations.

\begin{acknowledgments} We thank Ivan Pshenichnyuk and Andrzej Sobolewski for helpful and inspiring discussions. This work has been supported by the German-Israeli Foundation for Scientific Development (GIF) and the German Research Foundation (DFG) through the Cluster of Excellence 'Engineering of Advanced Materials' (EAM), SFB 953 and a research grant. Generous allocation of
computing time at the computing centers in Erlangen (RRZE) and Munich (LRZ) is greatly acknowledged.  
\end{acknowledgments}

\bibliography{bib,thoss,pbc}

\end{document}